\documentclass[preprint]{aastex} 

\newcommand{\beq}{\begin{equation}}
\newcommand{\eeq}{\end{equation}}

\newcommand{\kms}{\,km~s$^{-1}$ } 
\newcommand{\kmsii}{\,km~s$^{-1}$} 
\newcommand{\hmpc}{\,$h^{-1}$Mpc }
\newcommand{\hmpcii}{\,$h^{-1}$Mpc}
\newcommand{\speak}{\,\ensuremath{s_{\mathrm{p}}} }
\newcommand{\speakii}{\,\ensuremath{s_{\mathrm{p}}}}
\newcommand{\qvl}{\,DR7-Dim }
\newcommand{\vl}{\,DR7-Bright }
\newcommand{\vll}{\,DR7-Bright2 }
\newcommand{\qvlii}{\,DR7-Dim}
\newcommand{\vlii}{\,DR7-Bright}
\newcommand{\vllii}{\,DR7-Bright2}
\newcommand{\hgpc}{\,\ensuremath{h^{-3}\mathrm{Gpc}^3} }
\newcommand{\hgpcii}{\,\ensuremath{h^{-3}\mathrm{Gpc}^3}}
\newcommand{\baf}{\,baryonic acoustic feature }
\newcommand{\bafii}{\,baryonic acoustic feature}
\newcommand{\Epaper}{\,\cite{eisenstein05b} }
\newcommand{\Epaperii}{\,\cite{eisenstein05b}}

\newcommand{\DvR}{\,\ensuremath{1099}}
\newcommand{\dDvR}{\,\ensuremath{38}} 
\newcommand{\speakhR}{\,\ensuremath{103.6}} 
\newcommand{\dspeakhR}{\,\ensuremath{3.6}}  
\newcommand{\rsDvR}{\,\ensuremath{0.1394}} 
\newcommand{\drsDvR}{\,\ensuremath{0.0049}}  
\newcommand{\rsfid}{\,$r_s^{fid}(z_d)=159.75$ Mpc}
\newcommand{\nopeaks}{\,$10$}
\newcommand{\avg}[1]{\ensuremath{\langle{#1}\rangle}}

\begin{document}

\title{The Baryonic Acoustic Feature and Large-Scale Clustering \\ in the SDSS LRG Sample}
\author{Eyal A. Kazin,$^a$\footnote{eyalkazin@gmail.com} \ Michael R. Blanton,$^a$ Rom$\acute{\mathrm{a}}$n
Scoccimarro,$^a$ Cameron K. McBride,$^b$ \\ Andreas A. Berlind,$^b$ 
Neta A. Bahcall,$^c$ Jon Brinkmann,$^d$ Paul Czarapata,$^e$ 
Joshua A. Frieman,$^f$$^g$ Stephan M. Kent,$^h$ Donald P. Schneider,$^i$ Alexander S. Szalay$\ ^j$}

\affil{$^a$\footnotesize{Center for Cosmology and Particle Physics, New York University, 4 Washington Place, New York, NY 10003, USA}}
\affil{$^b$\footnotesize{Department of Physics and Astronomy, Vanderbilt University, 1807 Station B, Nashville, TN 37235, USA}}
\affil{$^c$\footnotesize{Department of Astrophysical Sciences, Princeton University, Princeton, NJ 08544, USA}}
\affil{$^d$\footnotesize{Apache Point Observatory, P.O. Box 59, Sunspot, NM 88349, USA}}
\affil{$^e$\footnotesize{Fermi National Accelerator Laboratory, P..O Box 500, Batavia, IL 60510, USA}}
\affil{$^f$\footnotesize{Particle Astrophysics Center, Fermilab, P.O. Box 500, Batavia, IL 60510, USA}}
\affil{$^g$\footnotesize{Kavli Institute for Cosmological Physics, Department of Astronomy \& Astrophysics, University of Chicago, 
Chicago, IL 60637, USA}}
\affil{$^h$\footnotesize{MS 127, Fermilab, P. O. Box 500, Batavia, IL 60510, USA}}
\affil{$^i$\footnotesize{Department of Astronomy and Astrophysics, The Pennsylvania State University, 525 Davey Laboratory, 
University Park, PA 16802, USA}}
\affil{$^j$\footnotesize{Department of Physics and Astronomy, The Johns Hopkins University, 3701 San Martin Drive, Baltimore, MD 
21218, USA}}

\begin{abstract}
We examine the correlation function $\xi$ of the Sloan Digital Sky
Survey (SDSS) Luminous Red Galaxy sample (LRG) at large scales
($60<s<400$\hmpcii) using the final data release (DR7; $105,831$ LRGs between $0.16<z<0.47$). 
Using mock catalogs, we demonstrate
that the observed baryonic acoustic peak and larger scale signal are
consistent with $\Lambda$CDM at the $1.5\sigma$ level.
The signal at $155<s<200$\hmpc tends to be high relative
to theoretical expectations; this slight deviation can be attributed to a
bright subsample of the LRGs.
Fitting data to
a non-linear, redshift-space, template based-model, we constrain the
peak position at
\speakii$=$\speakhR$^{+3.6}_{-2.4}$\hmpc when fitting the range
$60<s<150$\hmpc ($1\sigma$ uncertainties measured from the mocks).
This redshift-space distance \speak is related to the comoving sound
horizon scale
$r_s$ after taking into account matter clustering
non-linearities, redshift distortions and galaxy clustering bias.
Mock catalogs show that the
probability that a DR7-sized sample would not have an identifiable
peak is at least $\sim \nopeaks \%$.
As a consistency check of a fiducial cosmology, we use the observed
\speak to obtain
 the distance $D_V\equiv [(1+z)^2D_A^2cz/H(z)]^{\frac{1}{3}}$ relative
to the acoustic scale. We find
$r_s/D_V(z=0.278)=$\rsDvR$\pm$\drsDvR.
 This result is in excellent
agreement with \cite{percival09a}, who examine roughly the same data
set, but using the power spectrum.
Comparison with other determinations
in the literature are also in very good agreement.
We have tested our results against a battery of possible
systematic effects, finding all effects are smaller than
our estimated sample variance.
\end{abstract}


\section{Introduction}
The large scale clustering of matter is a critical test of any
cosmological model. $\Lambda$CDM models predict a matter
auto-correlation-function $\xi$ that crosses over from positive
correlations to negative correlations at scales of
$\sim140$\hmpcii. An interesting feature in $\xi$ is the baryonic
acoustic peak, a residual from plasma sound-waves that came to a near
stop at the end of the baryon drag epoch ($z_d\sim 1010$). The baryonic acoustic
signature is strongly imprinted in the Cosmic Microwave Background (CMB)
temperature fluctuations and observations determine its physical size
to high precision ($<1.3\%$; \citealt{spergel03a}). Predicted to appear
in galaxy clustering measurements, the baryonic acoustic feature can
in principle provide determination of cosmic distances with very small
statistical uncertainties ($<1\%$; \citealt{seo07}).

The feature in the galaxy $\xi$ was first measured by
\cite{eisenstein05b} using Luminous Red Galaxies (LRG) from the Sloan
Digital Sky Survey (SDSS; \citealt{york00a}). Using $\sim 47,000$ LRGs from the third
data release (DR3), they determined
the distance to $z\sim0.35$ at $5\%$ accuracy and constrained
cosmological parameters. 
Similar measurements using later data
releases detected a peak, but as we discuss below, a broader one
(\citealt{cabre09i}; \citealt{martinez08}; \citealt{labini09};
\citealt{sanchez09a}).  This apparent peculiarity may raise concerns about
the utility of the baryonic acoustic peak in distance determination
and the consistency of $\Lambda$CDM models in large-scale galaxy
clustering. 

\cite{padmanabhan07b} (DR3; \citealt{abazajian05b}) and \cite{blake07} (DR4; \citealt{adelman06a})
 showed in their
photo-z analysis of the power spectrum of the LRGs 
a hint of excess power at large
scales. \cite{blake07} suggest a few possible causes for this excess power: residual
systematic errors, cosmic variance, large-scale galaxy biasing
mechanisms, and new early Universe physics. \cite{cabre09i} used Data
Release 6 (DR6; \citealt{adelman08a}; $\sim 75,000$ LRGs, $\sim 1$\hgpc in comoving-volume)
to probe for possible systematics in data analysis, showing that the
strong large-scale signal at $s>130$\hmpc is persistent for various
choices of weighting schemes and galaxy sample selection.
\cite{martinez08} confirmed this stable, but wider \baf in
DR7 (\citealt{abazajian09a}). \cite{labini09} reanalyzed this last sample, pointing out that
there was no obvious cross-over to anti-correlations at the predicted
scale of $r_c \approx 140$\hmpc or even at $250$\hmpcii.

Other studies that investigate large-scale
clustering and the \baf of the SDSS LRGs and the 
2dF Galaxy Redshift Survey (\citealt{colless03a})
include \cite{cole05a}, \cite{tegmark06a}, \cite{percival07a},
\cite{sanchez09a}, \cite{percival09a}, and \cite{reid09a}.

The purpose of this study is to measure the LRG 
two-point correlation function on large scales, compare it 
with the predictions of the $\Lambda$CDM model, and derive 
constraints on the baryon acoustic oscillation scale. 
We obtain precise 
uncertainty estimates on $\xi$ by using a large suite of mock catalogs 
drawn from N-body simulations. In the course of our study, 
we examine the differences in $\xi$ at large scales among 
the different SDSS data releases; we show that 
the differences between the DR3 and DR7 (the latter contains
$\sim 105,000$ LRGs, $\sim 1.6$\hgpcii) results do not arise 
from known systematics in the data analysis.
Further, we demonstrate
that the stronger large-scale signal in the final data release
is consistent with the
$\Lambda$CDM framework. We also analyze the position of 
the \bafii, explaining systematics, and relate our measurement
to the cosmic distance $D_V$.

In \S\ref{sdssdata} we discuss the SDSS data set. In \S\ref{method} we
explain our methods and mock catalogs. In \S\ref{resultsdr3} we show
that we can reproduce results from \cite{eisenstein05b} using their
same sample (DR3). In \S\ref{testinglcdm} and \S\ref{brightsub} we apply
the same technique to calculate the large-scale $\xi$ of DR7 and
compare to LRG mock catalogs. In \S\ref{testinglcdm} we also investigate
 the chances of identifying a baryon acoustic peak in an SDSS-sized sample. 
In \S \ref{obssys} we test the robustness
of the \baf and large-scale signal in $\xi$ to systematic errors in
calibration, suggesting that SDSS photometric calibration errors
should not affect the final results. In \S\ref{position} we determine
the location of the peak of the \bafii, and its uncertainty. 
 In \S \ref{dvcompare} we use the peak position to determine the ratio
between the acoustic scale at the drag epoch $r_s$ to an effective distance $D_V$
at redshift of $z=0.278$. In the appendices we discuss technical aspects of the
selection function, systematic uncertainties and estimators.

In the following, unless otherwise indicated, all calculations assume
a flat $\Lambda$CDM model with present day matter density
$\Omega_{M0}=0.25$ and a present Hubble expansion rate $H_{0}=100 h$ 
\kms$\mathrm{Mpc}^{-1}$. When running mock simulations $h=0.7$,
but otherwise $h=1$ when converting redshifts to comoving distances.
 When analyzing the DR3 we use the same
cosmology as \cite{eisenstein05b}, $\Omega_{M0}=0.3$.

\section{Data}\label{sdssdata}

The SDSS is the largest volume LRG
survey to date, having imaged the sky at high Galactic latitude in
five passbands $u,g,r,i$ and $z$ (\citealt{fukugita96a},
\citealt{gunn98a}) using a $2.5$m telescope (\citealt{gunn05a}). The images are processed (\citealt{lupton01a},
\citealt{stoughton02a}, \citealt{pier03a}, \citealt{ivezic04a}) and
calibrated (\citealt{hogg01a}, \citealt{smith02a}, \citealt{tucker06a}), allowing selection
of galaxies, quasars (\citealt{richards02a}) and stars for spectroscopy
(\citealt{eisenstein01a, strauss02a}) with twin fiber fed double
spectographs. Targets are assigned to plug plates according to a
tiling algorithm ensuring nearly complete samples
(\citealt{blanton03a}).

The SDSS LRG sample developed by \citet{eisenstein01a} serves as a
good  tracer of matter as they are associated
with massive halos. The LRG high luminosity
enables us to obtain a large volume,
and their spectral uniformity make them 
relatively easy to identify. 


The SDSS LRG sample covers $19\%$ of the sky and
the sky distribution is shown in Figure \ref{aitoff}.
The sample includes a large quasi-volume-limited
region to a redshift of $z \sim 0.36$, and is flux-limited thereafter
extending to $z\sim 0.47$.
The peak at $z\sim 0.36$ is associated with the passage of the 4000\AA\ break into the
$r$-band.


From the full sample DR7-Full we subample to
answer two main purposes: Comparison
with previous studies, focus
on volume-limited regions.

To avoid the effects of the flux limitted region,
we focus much of our analysis on subsamples
of DR7 similar to those chosen by \cite{zehavi05a}.
 \qvl  ($0.16<z<0.36$, $-23.2<M_g<-21.2$) is produced by subsampling DR7-Full
at $z \leqslant 0.36$. This quasi-volume-limited subsample
is {\emph{not}}
a dimmer sample than DR7-Full, but rather is called "Dim"
as a moniker to distinguish from the overlapping brighter subsample
\vl ($0.16<z<0.44$, $-23.2<M_g<-22.8$).
For further tests this last sample was devided into two subsamples
\vlii-Near ($0.16<z<0.36$, $-23.2<M_g<-22.8$) and \vlii-Far ($0.36<z<0.44$, $-23.2<M_g<-22.8$).
 We also included results for another volume
limited subsample \vllii, which is a brighter subsample of \qvl
($0.16<z<0.36$, $-22.6<M_g<-21.6$).

All subsamples are summarized in Table 1. In addition to various cuts in
redshift and absolute magnitude $M_g$, we also analyze a sample limited to
the earlier release DR3 analyzed by \citet{eisenstein05a}.

Figure \ref{nofz} shows the comoving number density as a function of
redshift $n(z)$ for each sample. DR7-Full (black) is the full sample, from
which we subsample DR3 (cyan).
 DR3 covers only the sky area
 of DR7-Full which was also covered by DR3. For most calculations in DR3
we apply the fiducial 
cosmology chosen by \Epaperii. It yields a similar (but noisier) $n(z)$ and comoving 
volume density to the DR7-Full, differing somewhat due to the different
 cosmologies (flat $\Lambda$CDM, with $\Omega_{M0}=0.25$ in DR7 and
$\Omega_{M0}=0.3$ in DR3). Our subsample agrees in detail with
\Epaper\ with only $< 350$ mismatches out of $\sim 47,000$ galaxies
($0.7\%$). For a more thorough discussion of LRG selection please see
Appendix \ref{angular}. 
The figure also displays the radial selection function $n(z)$
of subsamples \qvl (green), \vl (blue) and \vll (red).

In Appendix \ref{radial} we show,
using mock catalogs, that features in $n(z)$ up to $z<0.36$ cause  a negligible
change in the noise properties of the sample relative to a
volume-limited one.

For reasons described in \S\ref{lasdamas} we only use the Northern
Galactic Cap for subsamples \qvlii, \vlii. For DR7-Full we make use of both
caps, as in DR3. As explained in Appendix \ref{various},
we verified that the resulting $\xi$ does not change
when excluding the Southern Galactic Cap region.

A physical constraint when obtaining
spectra, known as fiber-collisions,
is dealt with here when analyzing 
the data. Due to the physical size
of the fibers of the spectrometer,
one can not obtain spectra simultaniously
of two targeted objects that 
reside within $55''$ of each other.
To reduce this effect, regions in the sky
were overlapped by multiple observing plates.
We count $\sim 2 \%$
of targeted LRGs were missed due to
fiber-collisions. To account 
for this we up-weight LRGs
with spectra which are within $55''$ proximity
of LRGs without.
For more details of our method
for fiber-collision correction,
its impact on $\xi$ as well as
 acounting for holes
and boundary effects of the survey,
please refer to Appendix \ref{angular}. 

All data used in this study may be obtained on the World Wide Web.\footnote{http://cosmo.nyu.edu/$\sim$eak306/SDSS-LRG.html}

\section{Method}\label{method}

\subsection{Clustering Estimator and Random Points}

To measure the two-point correlation function we use the
\cite{landy93a} estimator:
\beq\label{xiab}
\xi(s)=\frac{DD(s)-2DR(s)+RR(s)}{RR(s)},
\eeq
which compares the normalized number of data pair counts to randomly
distributed points. The quantity $s$ refers to the mean redshift space separation
for each bin. If we define $r$ to be the ratio of the number of random
points to data and $N_{DD}(s)$ to be the total number of
galaxy pairs separated by values $(s-ds/2,s+ds/2]$ (where $ds$ is the
width of the bin), then the normalized number of pairs are
$DD=N_{DD}r^2$, $DR=N_{DR}r$, and $RR=N_{RR}$. Here, $DR$ and $RR$
stand for data-random, random-random pairs, respectively. The random
points account for the effects introduced by survey boundary, holes within the data set, and
sector incompleteness. 

Additionally, in our counting of pairs, we apply weights to each
galaxy.  Appendices \ref{masking} and \ref{schemes} discuss the
details of distributing the random points and the pair-count
weighting.  In Appendix \ref{various}, we also compare this estimator
to other known methods, showing excellent agreement with the method
proposed by \cite{hamilton93} on large scales, and substantial differences with those
proposed by \cite{davis83} and \cite{peebles74}.

\subsection {LRG Mock Catalogs}\label{lasdamas}

We use mock galaxy catalogs produced from the LasDamas simulations (McBride et al.~2009;
in prep) to measure the uncertainty covariance matrix, as well as to
investigate systematic errors in the two-point-correlation
estimators. These mock catalogs provide 160 light-cone redshift-space
realizations of an SDSS-sized volume, with appropriate number
densities and clustering properties for comparison to the observed
data.

The LasDamas simulations are designed to model the clustering of SDSS
DR7 in a wide luminosity range.  In this suite of simulations galaxies
are artificially placed in dark matter halos specifically using the
formalism of the halo occupation distribution 
(HOD; \citealt{berlind02a}) with parameters chosen to
match an observational SDSS sample.  For complete details see McBride
et al. (2009, in prep.); for distribution visit the the World Wide
Web.\footnote{http://lss.phy.vanderbilt.edu/lasdamas/}


We use the ``gamma'' release of mock LRG catalogs produced from 
simulations. The Oriana simulation consists of 40 $N$-body dark matter
simulations; each realization contains $1280^3$ particles of mass
$45.73 \ 10^{10}h^{-1}\mathrm{M_{\sun}}$ in a box of length
$2400$\hmpcii. The softening parameter is $53
h^{-1}\mathrm{kpc}$. The simulations assume a flat $\Lambda$CDM
cosmology with total matter density $\Omega_{M0}=0.25$, baryon density
$\Omega_{b0}=0.04$, Hubble expansion rate $H_{0}=70$\kmsii Mpc$^{-1}$,
$\sigma_8=0.8$, and spectral index $n_s=1$. The HOD parameters are
adjusted to reproduce the observed number density as well as the
projected two-point-correlation function $w_p(r_p)$ at projected
separations $0.3<r_p<30$\hmpcii, well below the scales we consider
here. The choice of HOD affects the resulting $\xi(s)$ shape on small
scales, but on large scales HOD primarily biases the amplitude as expected
from local galaxy bias arguments (\citealt{fry93,coles93,scherrer98,narayanan00a}).

The LasDamas team has divided the sky into two alternative footprints:
one can use either the SDSS Northern Galactic and Southern Galactic Caps
together, or only the Northern. We use the latter option because in that case
each simulation produces four samples (as opposed to two in the
former) resulting in twice as many realizations. In summary, we analyze
$40\times 4 =160$ LasDamas mock catalogs for each of both luminosity
subsamples \vl and \qvlii. We note that the angular distribution of
galaxies in these mocks is similar to the angular distribution of the
data, aside from the observational incompleteness. In Appendix
\ref{angular} we explain how we account for the incompleteness within the
observational data.

Our only manipulation of the catalogs provided by McBride et
al.~(2009; in prep) is in the radial direction, where we randomly
subsample to match the SDSS comoving volume density $n(z)$, to better
represent the Poisson noise. This subsampling reduces the number of
LRGs in \qvl by $15\%$ and in \vl by $7\%$. In Appendix \ref{radial}
we show that this subsampling does not noticeably
affect either $\xi$ or its uncertainties. 
For more details regarding the radial selection function,
please refer to that Appendix.

\subsection {Covariance Matrix and $\chi^2$ Fitting}\label{covmatrix}

In our analysis we use the standard $\chi^2$ fitting algorithm based on
a covariance matrix, which is constructed from the LasDamas mock
realizations. 

As the uncertainties in the $\xi$ values for the bins are
interdependent, we build a covariance matrix $C_{ij}$ to measure the
dependence of the $i^{th}$ bin on the $j^{th}$. We construct $C_{ij}$
from the individual mock realizations as follows:
\beq\label{covariance}
C_{ij} = \frac{1}{N_{\mathrm{mocks}}-1}\cdot 
\sum_{k=1}^{N_{\mathrm{mocks}}}\left(\overline{\xi}_i-\xi_i^k\right)
\left(\overline{\xi}_j-\xi_j^k \right),
\eeq
where $\overline{\xi}_m$ is the correlation value for the $m^{th}$ bin
of the mock mean and $\xi_m^k$ is the same for the $k^{th}$ mock
realization. In all calculations here, $N_{\mathrm{mocks}}=160$.

We can then estimate $\chi^2$ for our observational result
$\xi^{\mathrm{obs}}$ relative to models $\xi^{\mathrm{model}}$ using:
\beq\label{chi2}
\chi^2(\vec{\theta})=
\sum_{i,j=1}^{N_s} 
\left(\xi^{\mathrm{obs}}_{i}-\xi^{\mathrm{model}}_{i}(\vec{\theta})\right)
C_{ij}^{-1}
\left(\xi^{\mathrm{obs}}_{j}-\xi^{\mathrm{model}}_{j}(\vec{\theta})\right),
\eeq
where $\vec{\theta}$ are parameters of the model, and $N_s$ is the
number of separation bins used.

\section{Results}

Here we present our results and analysis of the redshift space angle-averaged
$\xi(s)$ of the SDSS LRGs in two releases: DR3 and DR7. In \S
\ref{resultsdr3} we show that we can reproduce the DR3 LRG selection
as well as the $\xi(s)$ results of \Epaperii. Differences between DR3
and DR7 are, therefore, not due to systematic differences in our
analysis from that of \Epaperii. In \S\ref{testinglcdm} we use mock
catalogs to show that the large scale clustering of DR7 LRGs is
consistent with the concordance $\Lambda$CDM model at the $1.5 \sigma$
level. Furthermore, in \S\ref{brightsub}, we investigate a bright
subsample, which has a strong large-scale signal relative to our
$\Lambda$CDM predictions, not seen in the other subsamples. In
\S\ref{obssys}, we propose a possible observational bias, and explain
the robustness of the \baf and large-scale clustering to this
effect. In \S\ref{position} we measure the peak position of the
\bafii. We use this last measurement in \S \ref{dvcompare} to 
determine the ratio $r_s/D_V$ and compare results to other 
studies.

\subsection{Reproducing the DR3 $\xi(s)$}\label{resultsdr3}

To demonstrate consistency with previous studies, we first compare our analysis of an
earlier data release, DR3, with that of \Epaperii. For this analysis
we create a DR3 LRG sample with the same criteria used by
\Epaperii, and a corresponding random sample. In this subsection we
assume $\Omega_{M0}=0.3$ in calculations.

In Figure \ref{dr3xis} we reproduce the redshift-space $\xi$ of the SDSS
LRGs first measured by \Epaperii. There is  good agreement between our
results (green diamonds) and theirs (red crosses). 
To investigate the minute remaining differences,
we also test using their random catalogs with our data set and
vice-versa. We conclude that the remaining subtle variations are due
to Poisson noise in the random catalogs.  As previously presented in
\Epaperii, we obtain a narrow peak at an apparent separation $s>100$\hmpc
(see \S \ref{position} for our analysis of peak position value), and a
steep slope that flattens at $\sim 135$\hmpcii.

The dashed lines in Figure \ref{dr3xis} display our result for the final
data release DR7 (sample DR7-Full), which runs through the same galaxy
selection algorithm and data analysis as DR3. The thick blue dashed line uses
the same cosmology as in DR3; for the thin black dashed line we alter
that cosmology to $\Omega_{M0}=0.25$, $\Omega_{\Lambda 0}=0.75$. The
binning in all cases is the same chosen by \Epaperii, with the
exception that for DR7 we extend the signal a bit further. In both DR7
cases the resulting position of the \baf is in fair agreement
(\S\ref{position}). However, we clearly see stronger power on large
scales, yielding a wider peak. In the next section we examine the
significance of this strong large-scale signal relative to our 
$\Lambda$CDM model predictions.


\subsection{Consistency of SDSS LRGs Clustering with $\Lambda$CDM}\label{testinglcdm}

We test the DR7 LRG clustering against a $\Lambda$CDM model by
comparing the observed $\xi$ to those yielded by the LasDamas mock LRG
catalogs (\S \ref{lasdamas}). We run the same analysis as before on
each subsample of DR7: \qvl and \vl (see Table 1 and Figure
\ref{nofz}). We also analyze DR7-Full, which has the disadvantage of
being flux limited at $z>0.36$ but can probe larger-scale modes. We find that \qvl is in clear
agreement with the model used in the simulations.  \vl also has 
a strong signal at $s>150$\hmpcii.

We first investigate the quasi-volume limited subsample \qvl showing
the SDSS results compared to the $160$ LasDamas mock realizations.
This subsample has a similar $\xi(s)$ to DR7-Full at most radii
$s<145$\hmpc (see Figure \ref{dr7subsamples} for direct comparison).
Our results are presented in Figure \ref{dr7qvl}.
The mock mean is indicated by the green solid line and the SDSS
results are the diamonds. The light gray shaded region represents the
area in which the $68.2\%$ of the realizations lie closest to the mean 
(that is, the $1\sigma$ uncertainties). The dark gray area is the same for
$95.4\%$ ($2\sigma$ uncertainties). The dotted blue lines
indicate the strongest and weakest signal of all mocks at each
separation (not one realization in particular).

The observed signal is clearly within the $1$--$2\sigma$ significance
level at all scales. To quantify the significance of the strong signal
at large scales ($s>130$\hmpcii) we reproduce this last result to
$s<400$\hmpc (top inset; larger separation binning). Although the
expected signal at the largest scales is very small relative to the
noise, in the top inset we include data as far out as possible
without significantly reducing pair counts due to edge effects.
For a discussion regarding edge effects, please refer to Appendix 
\ref{various}.

We also show in Figure \ref{dr7qvl} a histogram of the mean correlation on large scales
${\langle\xi\rangle}_{s=[130,400]h^{-1}\mathrm{Mpc}}$ (bottom inset;
red dashed line is SDSS, green dot-dashed is the mock mean), where the
brackets denote an average over all separation bins in the indicated
range. To further quantify the significance of the large-scale signal,
we measure the $\chi^2$ difference between the observed $\xi$ and the
mock mean using Equations \ref{covariance} and \ref{chi2}. In the last
equation we limit our bins to those between $130<s<400$\hmpcii.
Using the $10$ bins in the top inset (d.o.f$=10$) we measure a
normalized value of $\chi^2/\mathrm{d.o.f}=0.721$. We have tried several
definitions for the significance of large-scale power and all agree
that there is a $1.5 \sigma$ consistency with respect to the
$\Lambda$CDM plus HOD model used here.

Figure \ref{dr7dimmocks} shows several hand-picked mock realizations
to demonstrate the effects of variance. We caution the reader to avoid interpreting these
particular realizations as ``typical''. Instead, we have chosen
them to demonstrate the variety of signals we could be reasonably
expected from a survey the size of SDSS \qvlii. Whereas some realizations
(e.g., green-dotted and red-solid lines) have a similar clustering
signal to that observed (symbols), others do not show evidence of
the \baf (e.g. the orange dashed and cyan dot-dashed lines). 

Visually examining at all 160 realizations, we find that at least 17 ($\sim \nopeaks\%$) 
of the mocks have no peak at the expected $s\sim 100$\hmpcii. 
We examined by eye all 160 mock \qvl $\xi$ results and
defined a peak-less realization as one with no sign of a peak
within $95<s<120$\hmpcii. We took a liberal approach, so this result
should be considered a lower bound; i.e, realizations with subtle
peaks were counted as having a peak.  If we ask how many of the mocks
have very clear signs of a peak, we find about 75 out of the 160 do. 
This designation is subjective, of course. It is also difficult to compare these numbers 
with theoretical expectations, e.g. using random Gaussian field statistics is not 
enough as weak peaks in the Gaussian case 
(corresponding to the linear density field) are washed out by nonlinear 
motions.

To verify the result of feature-less realizations,
 we also checked the mock realizations in
another independent mock LRG catalog. In the Horizon-Run mock catalog
(\citealt{kim09a}), LRG positions are determined by identifying
physically self bound dark matter sub-halos that are not tidally
disrupted by larger structures. After adjusting their mock galaxy
catalogs to fit the SDSS radial function $n(z)$, we ran the same
analysis as for LasDamas and found that, of their 32 realizations, 5
showed no sign of the \bafii, comparable to the result obtained with
LasDamas mock catalogs.

\subsection{Bright Subsample \vl}\label{brightsub}

Figure \ref{dr7subsamples} also shows that the correlation function of
DR7-Full (dashed line) is stronger than \qvl (green diamonds) at
scales of $150<s<200$\hmpcii.  To help understand the difference
between these samples we examine another subsample, \vlii.

The correlation function $\xi$ of the brighter volume limited sample,
\vl (blue crosses), is stronger on all scales than DR7-Full and
\qvlii, which is expected, as bias is known to be a function of
luminosity (see e.g. \citealt{zehavi05a}). For example, Figure \ref{bias} shows
the relative redshift-space bias ratios, defined as
\begin{equation}
b \equiv \sqrt{\xi_{DR7-Bright}/\xi_{DR7-Dim}},
\end{equation}
showing that \vl is biased by a factor 1.15 on most scales relative to
\qvlii, in agreement with the mock catalogs (by design, of course, at
scales $r_p<30$ \hmpcii).

As discussed in \S\ref{position}, Figure \ref{dr7subsamples}
shows a good agreement in baryonic acoustic peak position among the
DR7-Full and \qvlii. However, the relative strengths of the large-scale
signal is worth investigating. Figure \ref{dr7vl} is similar to
Figure \ref{dr7qvl} for \vlii. The figure and its insets show that its
signal is significantly stronger than the mock values on scales
$s>130$\hmpcii.

The bottom inset of Figure \ref{dr7vl} is the histogram of
$\langle\xi\rangle_{s=[130,500]h^{-1}\mathrm{Mpc}}$ for the mocks,
compared to the SDSS value (red vertical dashed line).  We perform the
same $\chi^2$ comparison as in \qvlii, but this time for scales up to
$500$\hmpc (see Appendix \ref{various} for a justification of the larger
scale). This test yields $\chi^2/d.o.f=2.5$ for $11$ degrees of
freedom (separation bins of top inset), showing an unlikely fit with the
model used here.

We note that this result does not rule out all possible
$\Lambda$CDM and HOD models, because we have compared to only one
choice of parameters designed to fit statistics on smaller scales (see
\S\ref{lasdamas} for details).


We also split \vl into two subsamples: \vlii-Near ($0.16<z<0.36$; $\sim
16,500$ LRGs) and \vlii-Far ($0.36<z<0.44$; $\sim 13,800$ LRGs). 
Each produce noisy results for $\xi$, so we could not draw concrete
conclusions regarding large-scale clustering and the
\bafii. For example, we find, as shown in \cite{cabre09i}, that the
distant subsample \vlii-Far has no clear sign of a \bafii. In any case, both
subsamples showed strong clustering on large scales --- the signal in
the \vl subsample is thus not coming preferentially from high or low
redshift.

\subsection{Effects of Systematic Calibration Errors}\label{obssys}

Given the large scales and small signals we are probing here, we need
to test whether our results are sensitive to systematic errors in the
calibrations as a function of angle. \cite{eisenstein01a} cautions
that the number count of LRG targets is sensitive to small variations
in color cuts, especially in the $g$ and $i$ bands. Since the SDSS
targeting catalog is known to have calibration errors at the
$1\%$ level, the true color cuts applied vary across
the survey (\citealt{padmanabhan07a}). Such a variation might introduce an
artificial clustering signal in our analysis.

To test the possible effect of these errors on $\xi$, we subsample the
LasDamas \vl sample in a spatially varying fashion, and analyze the
correlation function of the resulting sample (using the original,
un-subsampled random catalogs). In detail, the subsampling factor
varies from $95$--$100\%$ in sinusoidal waves along the declination
direction with a wavelength of 10 deg. This choice is motivated by the
fact that the targeting catalog is separately calibrated in each stripe,
which spans 2.5 deg and are (very roughly) parallel to lines of
constant declination.

Comparing the resulting $\xi$ with the full mock catalogs, we find
insignificant effects for individual mock realizations and no
difference when averaging over all 160. We tested this both on the
full \vl and on a subsample \vlii-Far ($0.36<z<0.44$; see Table
1). Effects are significant only when enhancing the subsampling factor 
amplitude from $5\%$ to $15\%$, well above the expected systematic uncertainties due
to calibration. Even at this unrealistic uncertainty, the
\baf is noticeable, on average, though with a weaker amplitude. 

Notably, we cannot obtain the strong large-scale signal as observed
in Figure \ref{dr7vl} using any realistic level of calibration
uncertainty.

\subsection{Baryonic Acoustic Peak Position}\label{position}

Here we measure the baryonic acoustic peak position
\speakii. We first construct a model
$\xi^{model}$, which is related to the average correlation function
of the mock catalogs $\overline{\xi}$ by:
\beq
\xi^{model}(s)=\beta\cdot\overline{\xi}(\alpha \cdot s),
\eeq
where $\beta$ represents a bias term and $\alpha$ represents a change
in length scale.  We minimize $\chi^2$ over $\alpha$ and $\beta$ (see
Equations \ref{covariance} and \ref{chi2}), using the observed $\xi$
at scales near the peak. After $\beta$ and $\alpha$ are determined we
use a spline interpolation on the (very smooth) best-fit model to
pinpoint \speakii. This procedure is run on the observed $\xi$ and, as
explained below, for the individual mock realizations.

The advantage of using this approach over linear models is that
clustering nonlinearities, clustering bias and redshift distortions
are already taken into account in the simulations, as well as the
angular mask and radial selection. A disadvantage is that we are not
self-consistently altering the cosmology for each model.  However, 
in \S \ref{dvcompare}, when using \speak to determine cosmological
distances, we expect the uncertainties in assumptions we make to be small 
relative to the data uncertainties.

$\overline{\xi}$ used above is based on the average $\xi$
of 75 mocks with a clear peak (as is the case in the
observations). Another option we test is using $\overline{\xi}$
constructed from the mean of all 160 realizations. 
The 160 mock mean yields what peak position is expected given
a \qvl size sample, where the 75 mock model is the same
only given a realizaation with a clear peak.
As we demonstrate below, the choice of mock mean does not 
significantly vary results. 

In Figure \ref{clearvsall} we show the correlation functions of
$\overline{\xi}_{75}$ constructed from clear peaked realizations
(dashed bright green) and $\overline{\xi}_{160}$ constructed from
 the full catalog (solid black). As expected, both have 
roughly the same clustering at all scales, except that the latter
has a weaker peak. The postion of the \baf is nearly the same
($107.88$\hmpc for the former $107.96$\hmpc for the latter),
and they have roughly the same width.
The bottom panel shows the residual $\overline{\xi}_{75}-\overline{\xi}_{160}$.
As the observation has a clear peak we choose 
for our final result to use $\overline{\xi}_{75}$,
and also show that $\overline{\xi}_{160}$  yields a similar result.
For both we use the covariance matrix constructed from all
160 realizations.


Figure \ref{bafposition} diplays our result for \qvl using $\overline{\xi}_{75}$.
The upper left panel shows the distribution of the best-fit peak position for the
mocks and the observations. The vertical red line is the best-fit
for the observations, the vertical green line is for the mean mock
catalog. The upper right panel exhibits the distribution of $\chi^2$ per degree of
freedom. The vertical red line marks the results from the
observations and the histogram is that of 75 clear-peaked mocks.
  The bottom left panel is the normalized covariance matrix
$C_{ij}/\sqrt{C_{ii}C_{jj}}$. The bottom right panel compares the best-fit
model $\xi^{model}$ (blue solid line) to the observations (symbols), and  
the red arrow points to the fit peak value. 

The mean peak value for the mocks (red line on the main plot) is
$s_{\mathrm{peak}}^{\mathrm{mock}}=107.88^{+3.77}_{-2.51}$\hmpcii, where
the $1\sigma$ uncertainties are calculated from mock
realizations only with clear signs of a peak (about 75 out of the 160).
The individual mock \speak results are shown as the histogram
on the upper left plot and their normalized $\chi^2$ result in the
upper right plot. We obtain the uncertainty on the observations by
scaling these uncertainties for this mock catalog distribution by
$\speakii/s_{\mathrm{p}}^{mock}$.

Using the observed \qvl sample between $60<s<150$\hmpc we find the
peak to be at \speakii$=103.64^{+3.62}_{-2.41}$\hmpc
$(^{+3.5\%}_{-2.3\%})$ where these $1\sigma$ uncertainties are scaled
from our mock catalog results. For this fit, we find $\chi^2/{\mathrm
d.o.f.}=1.09$, where we used d.o.f.$=19$ (the number of data points
used minus two fitting parameters $\beta,\alpha$).  Thus, the fit is
excellent. By eye it appears poorer because it undershoots
the data at all points, as one's eye does not account for the
strong covariances among the bins (see lower left panel of Figure
\ref{bafposition}).

If we limit ourselves to the region $60<s<135$\hmpc we find $\speak=
105.96^{+3.70}_{-2.46}$\hmpcii. Although both results are
consistent, the dependence of \speak on fitting range indicates that
this sample is still limited in its power to constrain \speakii. 
We also changed the lower limit to various values between $[55,75]$\hmpcii,
constraining the upper bound to $150$\hmpc, and find $s_p$ does not 
vary more than $0.6$\hmpcii. If the lower limit is increased to $80$\hmpc
we obtain \speak$=102.01^{+3.56}_{-2.37}$\hmpcii, showing a little sensitivity
due to not using the full dip feature around $80$\hmpcii.

We perform the same analysis on $24$ jackknife subsamples of the
observed sample.
They are obtained by dividing an RA-Dec map into $24$
regions of same number LRGs and excluding one region each in turn. For
each jackknife subsample we calculate $\xi$ and run it through the
peak finder algorithm.
The uncertainties yielded from jackknifing yield $\sigma_{\pm}^{jk}=^{2.25}_{0.26}$\hmpcii, 
smaller than the sample variance indicated above. 

When using $\overline{\xi}_{160}$  we obtain for the data
between $60<s<150$\hmpc \speakii$=104.17^{+3.42}_{-3.05}$\hmpcii, very similar 
as before. The $\chi^2/$d.o.f is 1.13.

Besides sample variance, another source of uncertainty in \speak
 is due to random-shot noise.
In  Appendix \ref{randomnoise} we show that choices of different 
random catalogs (ratio of $r \sim 15.6$ random points per data) yield
slightly different \speakii. To reduce this effect we choose for
our final results to use $r\sim 50$.

We also test the effect of dilution on \speakii, i.e., choosing a
different cosmology when converting redshifts to comoving distances
(\citealt{padmanabhan08a}). In all above results we apply a flat
cosmology with $\Omega_{M0}=0.25$. When using a different value
$\Omega_{M0}=0.30$ (but maintaining mocks as before) we obtain
\speak$=102.70$\hmpc at $\chi^2/$d.o.f.$=0.96$.  As this result is well
within the $1\sigma$ variance and systematics explained above, we
conclude that the choice of cosmology does not change our results, but
may need to be considered when variance is reduced in future surveys.


We find that the $\overline{\xi}$ for the \vl subsample,
in the same binning used for \qvlii, is quite noisy and not useful
to measure \speakii. Although \vl covers a larger volume than
\qvlii, its density is over four times lower yielding a sample
with less than half the number of galaxies, severely increasing noise 
in our measurement.

We applied the same algorithm (same \qvl covariance matrix) on
DR7-Full and DR3 and find consistency with these results. A more
in-depth analysis would involve building a new covariance matrix for
DR7-Full, as it has a larger volume than \qvlii. We do not perform
this analysis, because the LasDamas mocks do not extend to that
volume. DR3 has fewer LRGs than \qvl resulting in a noisier signal, so
we do not find reason to do further analysis.

\subsection{Relating Peak Position \speak  to Cosmic Distance  $D_V$}\label{dvcompare}

As \cite{bashinsky01a} describe, the peak position \speak is closely
related to $r_s$, the physical comoving sound-wave radius at the end
of the baryon drag epoch. $r_s$ is determined to high precision in
physical units from CMB measurements (e.g., \citealt{komatsu09a}),
whereas we determine \speak in redshift space. Therefore, we can
compare the two to measure the relationship between redshift and
distance, as proposed by \citet{eisenstein99a}.

There are three important effects we must consider in such a
determination.  First, the relationship between the redshift space peak position \speak 
and $r_{s}$ is sensitive to non-linear clustering, redshift distortions and galaxy bias, the dominant effect being a
broadening of the peak (\citealt{eisenstein07b,crocce08,smith08,angulo08,seo08, eisenstein07, kim09a}) and a subdominant shift 
towards small scales  that is below current statistical uncertainties.
 
Second, the measured \speak is necessarily associated with a fiducial
cosmology used to interpret the redshifts and angular positions in
terms of comoving distances.  The fiducial cosmology we assume is the
same concordance flat cosmology as the mocks used
($[\Omega_{M0}^{fid},\Omega_{b0}^{fid},h^{fid}]=[0.25,0.04,0.7]$).
Technically this means that we use cosmological assumptions in
two steps of our algorithm: selection of LRGs by magnitude cuts,
and converting observed redshifts to comoving distances. For more details
please refer to Appendix \ref{various}

Third, we need to specify what ``distance'' we seek to measure, since
the definition of cosmological distance within the context of general
relativity is not unique. The relevant distance measures in this context are the angular diameter
distance $D_A(z)$ and the Hubble constant $H(z)$ at redshift $z$
(\citealt{hogg99cosm}). The former would be ideally constrained by
measuring \speak in a thin shell at radius $z$, and $H(z)$ by
measuring the line-of-sight clustering (\citealt{matsubara04}). These measurements
individually are hard to perform with the current data set (for
attempts in DR6 please refer to \citealt{okumura08},
\citealt{gaztanaga08iv}). Instead, we will constrain a combination of
the two following the standard approach in the literature (see Equation \ref{DvEQ} below).

To account for the above effects, the standard interpretation of the
acoustic peak in the correlation function is as follows (with an
analogous argument made for the power spectrum;
\citealt{percival07a}). First
we assume the proportionality:
\begin{equation}
\label{rsgsp}
r_s = \gamma s_p. 
\end{equation} 
Mock catalogs can be used to determine $\gamma$, as long as the
assumed cosmology is not far from the true one.  We have
done so here for the LasDamas mock catalogs, whose cosmological
parameters are well-motivated from the CMB and other constraints. We
calculate \rsfid \ for the mocks using Equation 1 from \cite{blake03},
and calculate the sound speed $c_s$ and the end of the drag epoch
$z_d=1012.13$ using Equations 4 and 5 from \cite{eisenstein98}. We
insert $\Theta_{2.7}=2.725/2.7$ as the temperature ratio of the CMB in
their Equation 5. Using $s^{mock}$ from \S\ref{position} and $h=0.7$ we obtain
$\gamma=1.037$.

Second, we construct the ``distance'' quantity:
\begin{equation}
\label{DvEQ}
D_V\equiv \left[(1+z)^2D_A^2cz/H(z)\right]^{\frac{1}{3}}. 
\end{equation}
This quantity is designed in such a way that the ratio
$D_V(\avg{z})/s_p(\avg{z})$ is approximately independent of the choice
of the fiducial cosmological model (\citealt{eisenstein05a,
percival07a, padmanabhan08a}). 

Third, we can rearrange these relationships to obtain:
\begin{equation}
\label{Dvtrue}
\frac{D_V(\avg{z})}{r_s} = 
\frac{D_V(\avg{z},\mathrm{fid})}{\gamma s_p(\avg{z})}
\end{equation}
where $s_p(\avg{z})$ is understood to be the inferred $s_p$ from
\S\ref{position} given the fiducial cosmology. Measuring 
$s_p$ thus yields the ratio of the distance at redshift $\avg{z}$ to
the acoustic scale. Given the acoustic scale $r_s$ in Mpc from the
CMB, we then can determine the distance $D_V(\avg{z})$ to that
redshift. As Equation \ref{DvEQ} shows, this distance measurement
constrains a combination of the angular diameter distance $D_A$ and
the Hubble constant $H(\avg{z})$.

Given the results in \S\ref{position}, we find
$r_s/D_V=$\rsDvR$\pm$\drsDvR \ at $\avg{z}=0.278$, with $1\sigma$
uncertainties of around 3.5\%. Both our values and uncertainty bars are in
excellent agreement with the analysis of the power spectrum of a
similar sample by \citet{percival09a}. Combining our result with the
sound-horizon $r_s=153.3$Mpc obtained from WMAP5 CMB (\citealt{komatsu09a}) we
find $D_V(0.278)=$\DvR$\pm$\dDvR \ Mpc.  

Figure \ref{Dvmeasure} presents our result compared with those
obtained by \cite{percival09a}, \Epaperii, as well as predictions of flat
$\Lambda$CDM cosmologies.  Our data point is the black diamond where
we choose to use the mean sample redshift $\langle z
\rangle=0.278$. The red and orange crosses are the values published in
\cite{percival09a} and \cite{percival07a}: $r_s/D_V(z=0.2)=0.1981 \pm
0.0071$, $r_s/D_V(0.275)=0.1390 \pm 0.0037$, $r_s/D_V(0.35)=0.1094
\pm 0.0040$ where the small red crosses results are indicated in Table
3 of \cite{percival09a}. 
The cyan cross is $r_s/D_V(0.35)=0.1097 \pm 0.0039$ 
obtained by \cite{reid09a} when analyzing the P($k$) of the
reconstructed halo density field of DR7 LRGs.
The blue square is the result obtained by
\Epaper ($D_V(0.35)=1370\pm64$ Mpc, if we use the CMB $r_s$ value as before).
We also add the result obtained by \cite{sanchez09a}
(purple triangle) who analyze the DR7 LRG $\xi$. Using only LRG clustering
they obtain $D_V(0.35)=1230\pm 220$ Mpc.
They show that when combining $\xi$ with CMB measurements
they obtain a tighter constraint $D_V(0.35)=1300 \pm 31$ Mpc.
We plot the latter where we use $r_s$ from CMB for plotting purposes. 

Our result is in perfect agreement with that
obtained by \cite{percival09a} at $z=0.275$.
Please keep in mind that both results, as well as those
obtained by other above studies that analyze the SDSS LRGs,
{\emph{are not}} independent, as we use
roughly the same data set. We are encouraged, nevertheless, that our
consistency check yields a result in agreement with \cite{percival09a}.

We comment that if we use the median redshift $z_{\mathrm{med}}=0.287$ 
rather than the mean $\avg{z}$ we obtain 
$r_s/D_V(z=0.287)=0.1354 \pm 0.0047$
 and $D_V(z_{\mathrm{med}})=1132\pm 40$ Mpc.


When comparing these results to cosmological predictions, we assume a
flat $\Lambda$CDM, and fix $\omega_M
\equiv\Omega_{M0}h^2=0.1358$. This constraint is motivated by its low
($2.7\%$) uncertainty in the WMAP5 CMB temperature measurements
(\citealt{komatsu09a}). $D_V$ depends both on $\Omega_{M0}$ and $h$
independently, thus its values vary. Meanwhile, $r_s$ depends on
$\omega_M$ (and $\Omega_{b}h^2$) so it is kept constant.  The
$1\sigma$ results of $r_s/D_V$ indicate that, 
constraining $w_M$ from CMB, the preferred region of the
matter density lies in the range $\Omega_{M0}=[0.25,0.31]$, and
$h=[0.66,0.73]$ in agreement with CMB and others. 
This is consistent for both choices of redshift (mean or median).
If we would plot median redshift result ($z=0.287$) in Figure \ref{Dvmeasure}
it would appear along the $\Omega_{M0}=0.28$ line.
We defer a full analysis of the cosmological implications.

Our $\xi$ results for DR7 LRGs as 
well as the covariance matrix 
 may be be obtained on the World Wide Web.\footnote{http://cosmo.nyu.edu/$\sim$eak306/BAF.html}

\section{Discussion}\label{discussion} 

The nature of the baryonic acoustic peak and larger scales in $\xi(s)$
have also been discussed in previous studies. \cite{cabre09i} used DR6
($77,000$ LRGs) and found a similar level of clustering to ours and
examined various possible data analysis systematic effects that might
cause the strong signal (the positive $\xi$ at scales larger than $150$\hmpcii).
 We followed similar steps with the addition
that we revisit DR3 to reproduce results from \Epaperii. In agreement
with \cite{cabre09i}, we show in our Appendix \ref{systematics} that
data analysis systematics have no significant effects on these results.

The main difference between DR3 and DR7 is the sky
coverage. DR7 covers over twice as much sky and, as opposed to the
DR3, is continuous. The latter should not be an important issue due to
the fact that the random points used in the $\xi$ estimator take into
account boundary effects and holes within the RA-Dec plane. The mock
catalog tests we conducted suggest that sample variance is a possible
explanation for the difference between the large-scale signals of the
two data sets.


\cite{martinez08} also present a wide \baf and large-scale clustering
in the $\xi$ for SDSS LRGs (DR7). It is worth noting that their definition of
$M_g$ is slightly different than the one used here and in
\Epaperii. Also, as described in Appendix \ref{angular}, we correct
for data angular incompleteness where they did not. Nevertheless, our
DR7-Full $\xi$ results are in fair agreement with their DR7-LRG.


 We show that sample variance affects not only the
shape of the signal at large scales (hence helping explain the broadness of the \bafii),
but also the probability of detecting a peak: 
we found that  at least $\nopeaks\%$ of the mock
realizations lack a baryonic acoustic signal. Nevertheless, we
show, in agreement with other studies mentioned here, that the SDSS
LRG sample contains a \baf which is stable within most 
redshift and $M_g$ cuts, as well as possible observational
bias. Larger surveys are underway to better measure this new holy
grail for cosmic distances. For example, the Baryon Oscillation Spectroscopic Survey
(BOSS) is estimated to map $1.5$ million LRGs in a 
much larger volume than the DR7,
up to $z\sim 0.7$ (\citealt{schlegel09a}).

We measure the observed peak position \speak to an accuracy of $\sim
3.5\%$ based on a model constructed from our mock catalogs results.
The main source of this uncertainty is due to sample variance,
of the \qvl subsample used.
Fitting data to  a redshift space, non-linear model,
we also explain sensitivity of determining the peak
position to the range of data points used, as well as shot-noise.
These systematics are shown to be less than $1\sigma$ of the 
sample variance, but should be considered when the latter is reduced.
We use our measurement of \speak to determine 
the ratio $r_s/D_V$ (\S\ref{dvcompare}).
Our result agrees very well with that obtained by  \cite{percival09a}
who analyze the oscillations in
the power spectrum and quote results at a very similar redshift. 
Note that we use shape information in the correlation function and do not 
marginalize over cosmological parameters, but rather tested consistency of one fiducial cosmology.
 However, our result does not have tighter 
constraints than that obtained by their study.
 This may be due to our mock catalog error bars being larger than the lognormal 
approximation used in \cite{percival09a}  or a 
difference in the range of scales used,
 among other things 
(see \citealt{sanchez08} for a comparison of
 the relative performance of $\xi$  and P($k$) estimates).

Regarding claims of the absence of anti-correlations at the largest scales 
(\citealt{labini09}; DR7), we point out that the mock realizations
show a large variety of crossover values $r_c$ from positive to
negative correlations. In Figure \ref{crossover} we show all 160 \qvl
mocks. Their crossover points are indicated by short green lines, and
that of the mean (white line) by the orange line at $\sim 140$\hmpcii. We
find $4\%$ of the mocks do not crossover before $200$\hmpcii, but this
value should not be taken too seriously as it increases with
wider binning, which causes less noisy results and less crossovers. 
A similar result was showed to us by E. Gazta\~naga for DR6 mock catalogs.
The $r_c$ values (or $s_c$ as we measure in redshift space) are summarized in the histogram in the inset, showing a
wide variety of crossovers between $[120,160]$\hmpc and even some
around $80$\hmpcii. We comment that the crossovers are defined as the
first time the $\xi$ crosses through the zero value, and we do not
account for returns to positive values.
Though having different bias in
clustering in respect to matter, 
galaxies should have the same crossover point
between correlation and anti-correlation.
We conclude that sample uncertainties still dominate our 
ability to perform such a test for determining $r_c$.



\section{Summary and Conclusions}

Data sets later than DR3 yield a broader baryonic acoustic peak and stronger 
large-scale clustering signal than measured by \Epaperii. 
In this paper we have demonstrated that:
\begin{enumerate}
\item Differences between DR3 and DR7 results are not due to known
systematic uncertainties in data analysis.  Applying the same methods in
the DR7 analysis, we reproduce the same DR3 data set as \Epaper and
match the same resulting $\xi$. 
\item Large-scale clustering of DR7 results are in agreement with the
$\Lambda$CDM+HOD model used here (flat $\Omega_{M0}=0.25$). The
average $\xi$ at scales $130<s<400$\hmpc is within $1.5\sigma$
variation of mock LRG catalogs.   
\item{} The detected baryonic acoustic peak position in DR7 seems
stable within most investigated subsamples and agrees with
\Epaperii. When analyzing $\xi$ results between $60<s<150$\hmpc we
find the peak position to be at
\speakii$=$\speakhR$\pm$\dspeakhR\hmpcii. This result is sensitive up
to $1\sigma$ to the upper boundary of range chosen for analysis, 
random-shot noise and binning.
\item{}Using this last result we measure the ratio between the
sound-horizon at the end of the baryon drag epoch $r_s$ to $D_V$ (Equation
\ref{DvEQ}) at $r_s/D_V(z=0.278)=$\rsDvR$\pm$\drsDvR \ Mpc. Our result is
in excellent agreement with \cite{percival09a}, who analyzes the power
spectrum of roughly the
same sample, and utilizing a different approach
of analysis. Inserting $r_s$ obtained from WMAP5 we calculate
$D_V(0.278)=$\DvR$\pm$\dDvR \hmpcii.
Comparison with other determinations in the literature are also in very good agreement (see Figure \ref{Dvmeasure}). 
\item{}We find a lower bound of $\nopeaks\%$ of mock realizations that do
not show evidence of a sign of a \bafii. However, given a mock realization with a
clear peak, we conclude that we can measure its peak position
value \speak within $1\sigma_{s_{\mathrm{p}}}\sim 3.5 $\hmpcii. 
\item A bright volume-limited subsample of DR7, \vlii, shows a
significantly stronger large-scale-signal than predicted by our mock catalogs,
which is not explained by potential systematics.
\end{enumerate}

The differences between DR3 and DR7 in the $\xi$ may be explained in
two ways: signal variance or observational systematics. Our analysis
of the LasDamas mock catalogs show that the signal is still dominated
by noise, which yields a variety of large-scale signals, wide and
narrow baryonic acoustic peaks, as well as some featureless.

In \S\ref{obssys} we test a method of how the sensitivity of LRG
selection might affect the correlation function. Although \cite{eisenstein01a} cautions
that the number count of LRG targets is sensitive to small variations in 
color cuts, our test shows robustness of the large-scale clustering, and of the \baf in particular. 
\cite{hogg05a} also demonstrates consistency of the survey calibration in its different patches,
by examining DR3 SDSS LRG number counts in different regions of the sky.

Our analysis of the apparent strong large-scale signal 
of the \vl subsample compared to the model used does not 
have the power to rule out clustering predicted by $\Lambda$CDM,
as it is tested against one cosmology, and one HOD model.
As explained, the HOD parameters, 
were fit using only small-scale clustering, and might not be the best choice 
on large scales. However, it is reassuring that even in this case, 
the discrepancies of the data compared to our mocks are not much larger than
at the $2\sigma$-level (in order to rule out the cosmology used, one would have to marginilize
over all possible HOD fits, which we have not done).
 Furthermore, a large-scale enhancement of the two-point correlation 
function for a fixed cosmology may be obtained in models of primordial
 non-Gaussianity with parameter $f_{\rm NL}>0$, see e.g. Fig.~10 in \citet{desjacques09}.
It will be interesting to follow up  these issues 
with future surveys that will tighten uncertainties.


In summary, we show in this study that the SDSS LRG DR7 sample is
consistent with $\Lambda$CDM, and the \baf is stable within its various
datasets and most subsamples. 
 Funding for the SDSS and SDSS-II has been provided by the Alfred P. Sloan Foundation, the Participating Institutions, the National Science Foundation, the U.S. Department of Energy, the National Aeronautics and Space Administration, the Japanese Monbukagakusho, the Max Planck Society, and the Higher Education Funding Council for England. The SDSS Web Site is http://www.sdss.org/.

 The SDSS is managed by the Astrophysical Research Consortium for the Participating Institutions. The Participating Institutions are the American Museum of Natural History, Astrophysical Institute Potsdam, University of Basel, University of Cambridge, Case Western Reserve University, University of Chicago, Drexel University, Fermilab, the Institute for Advanced Study, the Japan Participation Group, Johns Hopkins University, the Joint Institute for Nuclear Astrophysics, the Kavli Institute for Particle Astrophysics and Cosmology, the Korean Scientist Group, the Chinese Academy of Sciences (LAMOST), Los Alamos National Laboratory, the Max-Planck-Institute for Astronomy (MPIA), the Max-Planck-Institute for Astrophysics (MPA), New Mexico State University, Ohio State University, University of Pittsburgh, University of Portsmouth, Princeton University, the United States Naval Observatory, and the University of Washington.
    
Fermilab is operated by Fermi Research Alliance, LLC under Contract No. DE-AC02-07CH11359 with the United States Department of 
Energy.	

It is a pleasure to thank Ariel  S$\acute{\mathrm{a}}$nchez for
his detailed comments on our manuscript.
We thank Daniel Eisenstein for assistance in selecting LRGs and 
Idit Zehavi for discussions on weighting algorithms and selection function considerations.  
We also thank Raul Angulo, Enrique Gazta\~naga, David Hogg, 
Will Percival, Ramin Skibba and Martin White 
for useful discussions and insight. 
We thank the LasDamas project for making their mock catalogs publicly available
We thank Horizon team for making their mocks public,
and in particular  Changbom Park and Juhan Kim for discusions on usage.
E.K thanks Adi Nusser for his hospitality at the Technion, Israel Institute of Technology,
and Betty K. Rosenbaum for her comments.
E.K was partially supported by a Google Research Award.
M.B was supported by Spitzer G05-AR-50443, NASA Award NNX09AC85G.
R.S. was partially supported by NSF AST-0607747 and NASA NNG06GH21G.    
    
\clearpage

\appendix{}
\section{Selection Functions}\label{masking}

Here we explain the angular and radial selection functions of the SDSS
LRG sample, as well as the mock catalogs. When selecting and weighting
LRGs in DR3 and DR7, we followed the procedures described in
\cite{zehavi05a} for the most part, and explain here a few differences
from that study, none of which affect the end results.


\subsection{Angular Mask}\label{angular}

The SDSS LRGs cover roughly $20\%$ of the sky, mostly in the Northern
Galactic Cap, and partially in the Southern (Figure
\ref{aitoff}). After an image of the sky is obtained in the
$u,g,r,i,z$ bands, objects are targeted for a spectroscopic
observation to measure their redshifts. Here we refer to targeted
quasars, Main galaxies and LRGs as {\emph{targeted object}}s.
\cite{zehavi05a}, in their appendix, discuss reasons that some
targeted objects fail to receive spectra, and their method (which is
also used in \citealt{eisenstein05b}) of correcting for this small
incompleteness. Please refer to their study for definitions of
``sector'', ``sector completeness'', ``fiber-collisions'' and the
polygon method with which the complicated angular geometry is
expressed in DR3 and later releases.

We choose a slightly different definition for sector completeness than
\cite{zehavi05a}. We define it as the number of targeted objects
(Quasars+LRG+Main) that have obtained spectra (after fiber-collision
corrections) divided by the total number of targeted objects.
By ``fiber-collision corrections'' in the number of targeted objects
which obtained spectra, we mean that before calculating sector
completeness, we up-weight objects with spectra which are within $55''$
of objects that did not. For example, if we find a group of
4 objects in $55''$ proximity and three of them get spectra, all three
are fiber-collision weighted by $4/3$. This up-weighting plays two
roles in our algorithm: defining sector completeness and while
counting galaxy pairs.

We mentioned above our definition for sector completeness. If, for
example, the only objects which did not get spectra within a sector are
due to fiber-collisions, the sector is considered fully complete. A
rare peculiarity occurs due to some $55''$ neighbors at different
sides of a border between sectors in which one (or more) obtain
spectra and another does not. In these cases some sectors might have a
completeness larger than unity, meaning they obtain a partial
completeness fraction from the neighboring sector. In summary, the
average completeness of all sectors is $98 \%$ for DR7. If we define
Main+LRG as 'targeted objects' (excluding quasars) the completeness
yields the same, and if we define targeted objects as LRGs only we
obtain $96 \%$. We find that these different definitions for
completeness result in subtle mismatches of the DR3 LRGs compared
sample to \Epaperii, all of which less than $0.7\%$ of the $\sim 47,000$
galaxies, and negligible effects on $\xi$ compared to shot-noise of
random points used.  As in \Epaper we limit ourselves to sectors with
$>60\%$ completeness. This results in 29 sectors (that have a non-zero
completeness) with a total area of $13$ square degrees ($0.16 \%$ of
targeted sky) and a total of $364$ targeted LRGs. 
As in \Epaper we used the completeness as
a probability to exclude random points from each sector. 
They also up-weight both data and random points
in each sector by the reciprocal of the completeness value.
With the high
completeness of the survey we do not expect differences in the
resulting $\xi$ to both methods.

To account for fiber-collisions while pair-counting, we use a slightly
different population than before. LRGs get up-weighted if they are
$55''$ neighbors of a targeted LRG (not quasar or Main galaxy) which
did not get spectra. This effectively increases our sample LRG number
by $\sim 2.2\%$ for DR7-Full and $\sim 1.8\%$ in DR3, which is
important to take into account when calculating normalized number 
of pairs DD, DR and RR.

The LasDamas mock catalogs match the survey geometry as described by
the polygon description in the NYU-VAGC.  Since the "gamma" release
mocks do not model fiber collisions nor missing sectors, completeness
is defined to be 100
 For more details please refer to McBride et. al (2009; in prep).

\subsection{Radial Selection Function}\label{radial}

The LRG sample used here is the largest quasi-volume-limited
spectroscopic sample of its kind today. That said, the radial
selection function, $n(z)$ is not constant (meaning not volume
limited), as one would expect in a homogeneous universe (up to Poisson
noise and radial clustering). Instead it is quasi-volume-limited up to
$z\sim 0.36$ and flux limited thereafter. In what follows we show that
the features in the $n(z)$ of
\qvl do not affect our results. On a more technical note we discuss
distributing random points in the radial direction.


We test the LasDamas mock catalogs for the features in the observed
radial selection function. The original mocks do not trace the $n(z)$
of the corresponding SDSS redshift and luminosity cuts. The top left
inset in  Figure \ref{xisfeffectplot} shows this difference in shape for the \qvl sample,
as LasDamas provides an $n(z)$ with a slight negative slope (cyan;
average over 160 realizations).  
This slope in the LasDamas mocks is a
consequence of applying fixed HOD parameters to a light-cone halo
catalog, as it neglects the evolution of the dark matter halo mass
(i.e. lower halo masses as higher redshift result in fewer artificial
galaxies and hence the negative slope).
The survey result (thick green histogram) is the same as in
Figure \ref{nofz}. Two features noticeable are the negative
slope between $0.16<z<0.28$, and the positive slope to the peak at $z \sim
0.34$. We show that these features have a negligible effect on $\xi$
and the r.m.s $\sigma_\xi$ compared to the distribution given by the
original mocks, which is closer to volume limited. To validate this
claim, we exclude mock LRGs such that they match the SDSS radial
selection function. For the \qvl this meant excluding $15\%$ of the
mock galaxies ($n(z)$ average over 160 realization shown as black histogram)
 and for \vl it meant excluding $7\%$, yielding, on
average, a similar number count and volume density as in Table 1. In
the main plot of Figure \ref{xisfeffectplot} we compare results before and after
exclusion of the mock galaxies, and find that the mean $\xi$ of 160
realizations agrees on all scales. The right inset shows that the diagonal
terms of the covariance matrix $\sigma_\xi \equiv \sqrt{C_{ii}}$ is
slightly higher for the galaxy-excluded sample as expected from
slightly larger Poisson noise. We conclude that the shape of observed
$n(z)$ for the \qvl sample does not appear to affect the results
obtained in this study.

We would also like to address the issue of distributing redshifts
to random catalogs, a non-trivial step in using random-point based
$\xi$ estimators, when dealing with non-volume-limited data. Two
popular methods for distributing redshifts (or comoving distances) to
random points are redistributing the actual data redshifts randomly,
and assigning random distances, so that the overall $n(z)$ shape
matches that of the data (i.e, using the data $n(z)$ as a probability
function). We test both random point distribution with and without
radial weighting.

 We show results of both methods of 
distributing random points in Figure \ref{randdistribute}.
To clarify differences, in the top panel we plot $s \cdot \xi$ and
the ratio between all cases to Data-Redshift 
in the bottom panel.
 For convenience we define the case of 
distributing data redshifts to random points as ``Data-Redshift'' (diamonds),
and the case of distributing random points to match the data $n(z)$
as ``Data-$n(z)$'' (crosses). The data used in this analysis are the original
\qvl mocks averaged over 8 realizations. We compare results without 
radial weighting (black) as well as with (bright green, shifted by 2\hmpc for visual 
clarity, shift not calculated within $s \cdot \xi$).
 The radial weighting
when pair counting is explained bellow. The figure shows
that ``Data-Redshift'' is noticeably weaker than ``Data-$n(z)$''
until $s\lesssim 110$\hmpcii. This is a clear display of the 
diminishing of the radial clustering modes.
Although this effect is very small relative to our current
 $1\sigma$ measurements, it will be important once 
these are reduced.
On the larger scales the differences are diminished. The red 
crosses emphasize the importance of weighting the random
points in the same fashion as the data. Not doing so
yields spurious clustering.

In conclusion, we choose to radially distribute random points to 
match the same $n(z)$ shape of the data, using the same weight,
and {\it{not}} distributing the data redshifts to the random points,
as to preserve the radial modes.


After deciding to use the survey $n(z)$ to  analyze the mock catalogs we have 
two options of use of random points:
(a) compare all data points to one random set, which has the mean $n(z)$ of the data;
or (b) imitate the observation for each realization by comparing each to a random 
catalog with a tailor-made random catalog. By ``tailor-made'' we refer to the adjustment 
of the redshifts $z$, so that each random catalog yields a similar $n(z)$ to that of
 the mock data. We applied this difference when analyzing \qvl and \vl and find no 
difference in the $\xi$ mean results. As for the r.m.s $\sigma_\xi$ we find very small
 differences in both. On scales up to $s<200$\hmpc the \qvl shows fluctuations smaller 
than $4\%$ where \vl shows fluctuations below $8\%$, with no particular preference
for either method. This might be a result of the fact that the transverse modes dominate
the radial ones. For completeness we note that
 in our mock results we use the former method, i.e, we apply one random catalog for all the mock data,
where the ratio of random to data is $r \sim 10.5$.

To weight the data approximately according to volume, we need to
account for the radial selection function $n(z)$ (see Appendix
\ref{radial} and Figure \ref{nofz}). To do so, we apply the standard
weighting technique (\citealt{feldman94}).

For the SDSS LRGs we calculate $n(z)$ in bins of $\Delta z=0.015$ and
use a spline interpolation to calculate the radial selection value of
each LRG and random point. Each point is then assigned a radial weight
of $1/(1+n(z)\cdot P_w)$ where we use $P_w=4\cdot 10^4
h^{-3}\mathrm{Mpc}^{3}$, as in \Epaperii. The weights are applied to
both the data and the random catalogs. The choice for $\Delta z$ does
not affect the measured $\xi$ (as also shown in
\citealt{cabre09i}). We also examined for differences between choosing the
observed $n(z)$ for weighting and the model used in \cite{zehavi05a}
and found no difference.

\section{Systematics of $\xi(s)$}\label{systematics}

Here we address possible data analysis effects on our results. We
focus on weighting schemes both in the radial and in the angular
masks. In general, we find that the position and width of the \baf does
not change much, though we do find small effects in \speak
due to Poisson shot-noise.

\subsection{Random Shot Noise}\label{randomnoise}

When comparing between different $\xi$ results obtained by various
systematics, it is important to differentiate biases from variances 
obtained by random shot-noise. 

Applying five different random catalogs to the observed \qvl,
at a ratio of $\sim 15.6$ random points for every data point, and 
assigning radial weighting (similar to 
that performed in the analysis) we show in Figure \ref{diffrandcat}
that the random shot-noise is minimal on most scales of concern. The
top panel shows the $\xi$ for five different random catalogs against 
the same observed sample (black-dotted, green-dashed, red-dot-short-dashed,
orange-long-dashed, blue-triple-dotted-dashed). To facilitate noticing
differences the bottom panel shows $\xi$ ratios in respect to the first
random catalog. For the chosen binning, the baryonic acoustic region seems
 to yield less than a $10\%$ difference. We reran our \speak algorithm
 on all random catalogs and find the peak positions 
 $[102.3,102.9,103.0,105.3]$\hmpcii. 
This shows that, although the overall shape seems very consistent when
using the different random catalogs, the random-shot noise
has a small effect on pinpointing peak position. This is currently 
smaller than the survey $1\sigma$, but should be considered when statistical uncertainties improve. 
The fitting normalized $\chi^2$ range for all five catalogs is 
$\chi^2/d.o.f=[1.08-1.38]$.
To reduce this effect on our measurement in \S\ref{position} we 
used a ratio of number of random to data of $\sim 50$.
The peak position for is marked by the top arrow
at \speak$=\speakhR$\hmpcii.

In the region between $50<s<90$\hmpc we see up to $10\%$
differences, which is expected in a sharp logarithmic slope. On smaller scales
$s<50$\hmpc differences yield up to $3\%$ difference in amplitude.

\subsection{Effects of Weighting}\label{schemes}

In order to optimally measure the correlation function and to account
for the fiber collision effects, we apply weighting algorithms. All
differences due to choices about how we weight turn out to be much
smaller than current $1\sigma$ variances. Angular weighting schemes
(fiber-collision correction and sector completeness) are explained in
Appendix \ref{angular}. Here we explain our algorithm for the radial
weighting and show results for both.

Figure \ref{weightplot} shows the effects of various angular and
radial weighting schemes on the large-scale $\xi$ of DR7-Full (for
clarity we present $s \cdot \xi$ in the top panel).
 The weighting schemes compared are:
\begin{enumerate}
\item No weighting at all (black)  
\item Radial weighting only (cyan; shifted by $1.75$\hmpc for clarity)
\item Radial weighting + Fiber-collision (red; shifted by $3.5$\hmpc)  
\end{enumerate}
As mentioned in Appendix \ref{angular}, sector completeness is taken
into account in the distribution of the random points.  
The bottom panel shows the ratio of each of the above options $\xi_i$
over the non-weighted $\xi_0$.
To avoid possible random shot noise (see Appendix \S \ref{randomnoise}) we use
the same set of random points in each case. The weight is modified for
each option for the data and random as indicated in the plot.  The
fiber-collision weight for the random set is always $w=1$.

We clearly see from the three above options the effect of radial
weighting on scales $s\lesssim 170$\hmpcii. We can grossly divide this
range into a smaller one of $s\lesssim 115$\hmpc in which the radial
weight adds some power to the signal, and a larger scale region in
which the signal is reduced. The fiber-collision correction weights
(``Fiber-Coll'') are important when dealing with scales $s\lesssim
40$\hmpc to the few percent level, but not at ranges discussed in this work.  Beyond $170$\hmpc
the effects of the weighting schemes used here are minimal. Most
importantly, as shown in \cite{cabre09i}, the apparent strong $\xi$
signal at large-scales, and the \baf position are consistent among the
various weighting methods.

These weighting effects are not substantial in Figure
\ref{randdistribute}, because the $n(z)$ used there (full \qvl LasDamas
catalog without dilution to match SDSS $n(z)$) does not have the same
complex features seen in DR7-Full (the mocks used in that figure are
instead close to volume-limited).

\subsection{Various Systematics}\label{various}

We have checked a few other possible systematics 
collectively described here.

\begin{figure}[htp]
\begin{center}
\includegraphics[width=\textwidth]{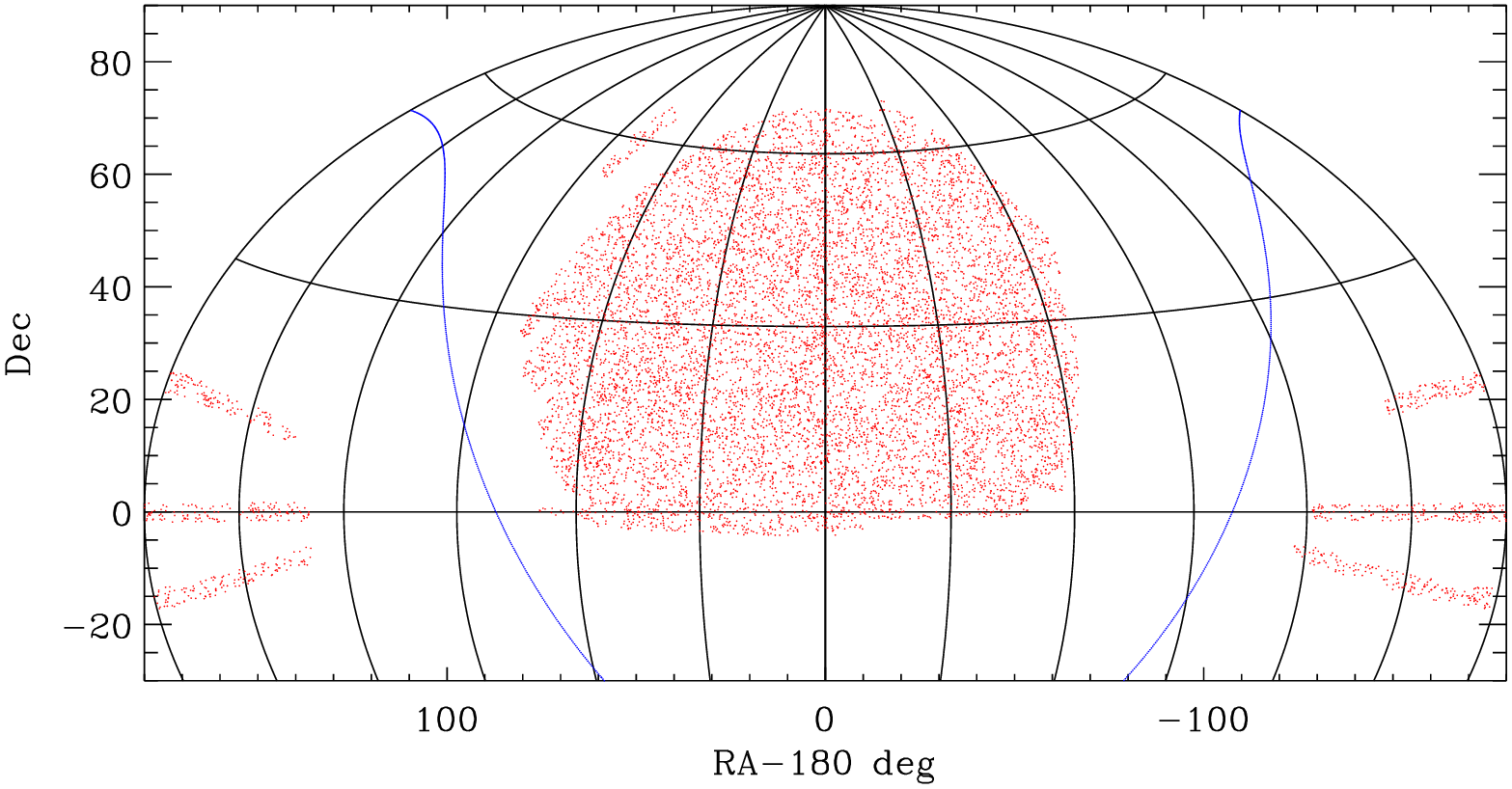}
\vspace{0.5in}
\caption{Angular Selection Function : SDSS DR7 LRG sky coverage. For
plotting purposes we present one tenth of the $105,831$ DR7-Full
galaxies. The solid blue line is the Galactic plane. }
\label{aitoff}
\end{center}
\end{figure}

\begin{figure}[htp]
\begin{center}
\includegraphics[width=\textwidth]{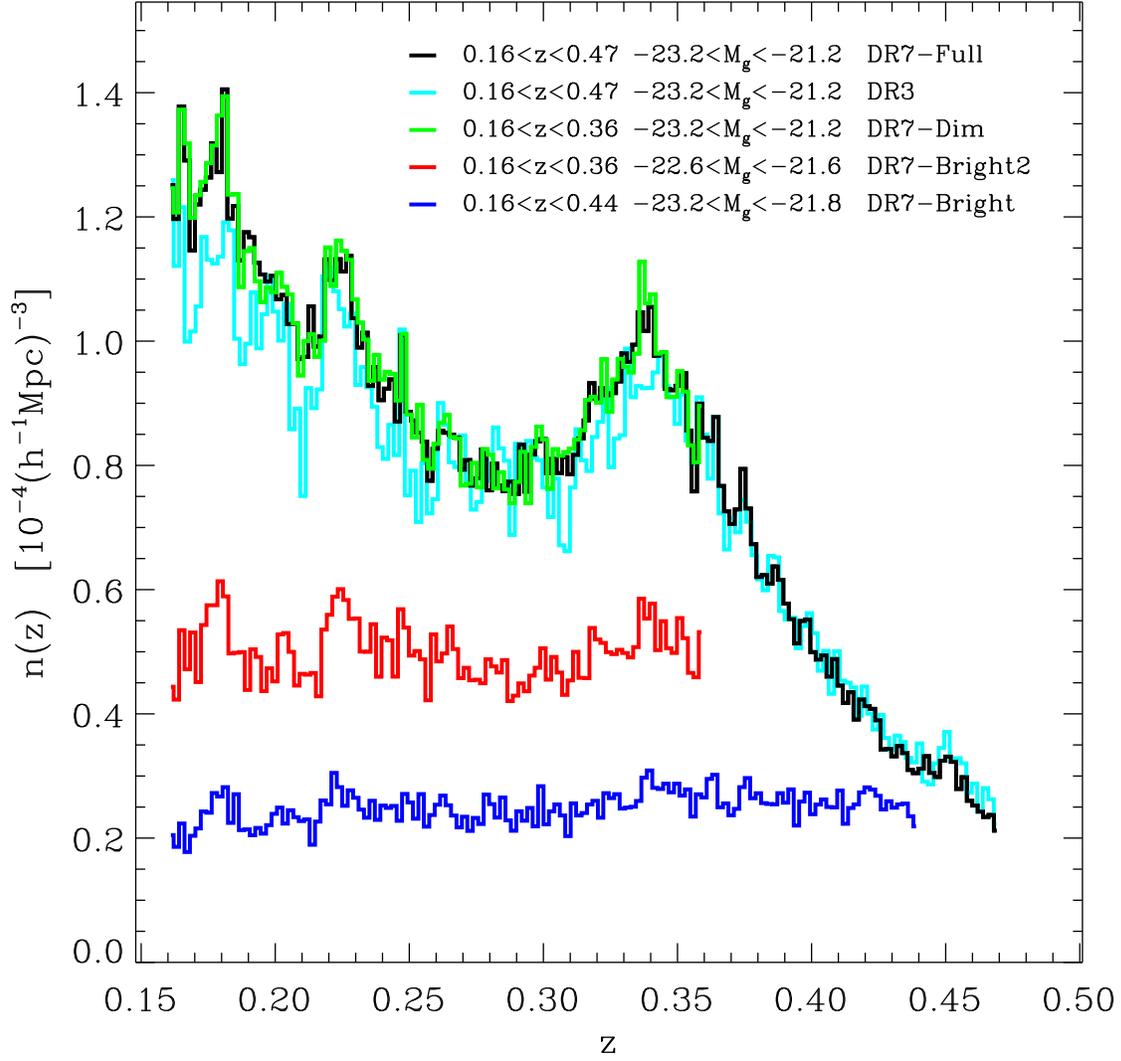}
\caption{Radial selection function: Comoving number density $n(z)$ of
the full DR7 (DR7-Full; black) 
and its subsamples \qvl (green), 
\vl (blue), \vll  (red)
and DR3 (cyan).
} 
\label{nofz}
\end{center}
\end{figure}

The SDSS DR7 sample is mostly contiguous, with only $9.8\%$ of the
surveyed area not contained in the main part of the Northern Galactic
Cap sample (Figure \ref{aitoff}). We examined three choices for which
footprint to use: the full survey ($100 \%$ of LRG sample), the
Northern Galactic Cap only ($90.9\%$), and the Northern Galactic Cap
without the small ``island'' ($90.2\%$).  We found no significant
difference in the resulting DR7 $\xi$ for these samples.

\begin{figure}[htp]
\begin{center}
\includegraphics[width=\textwidth]{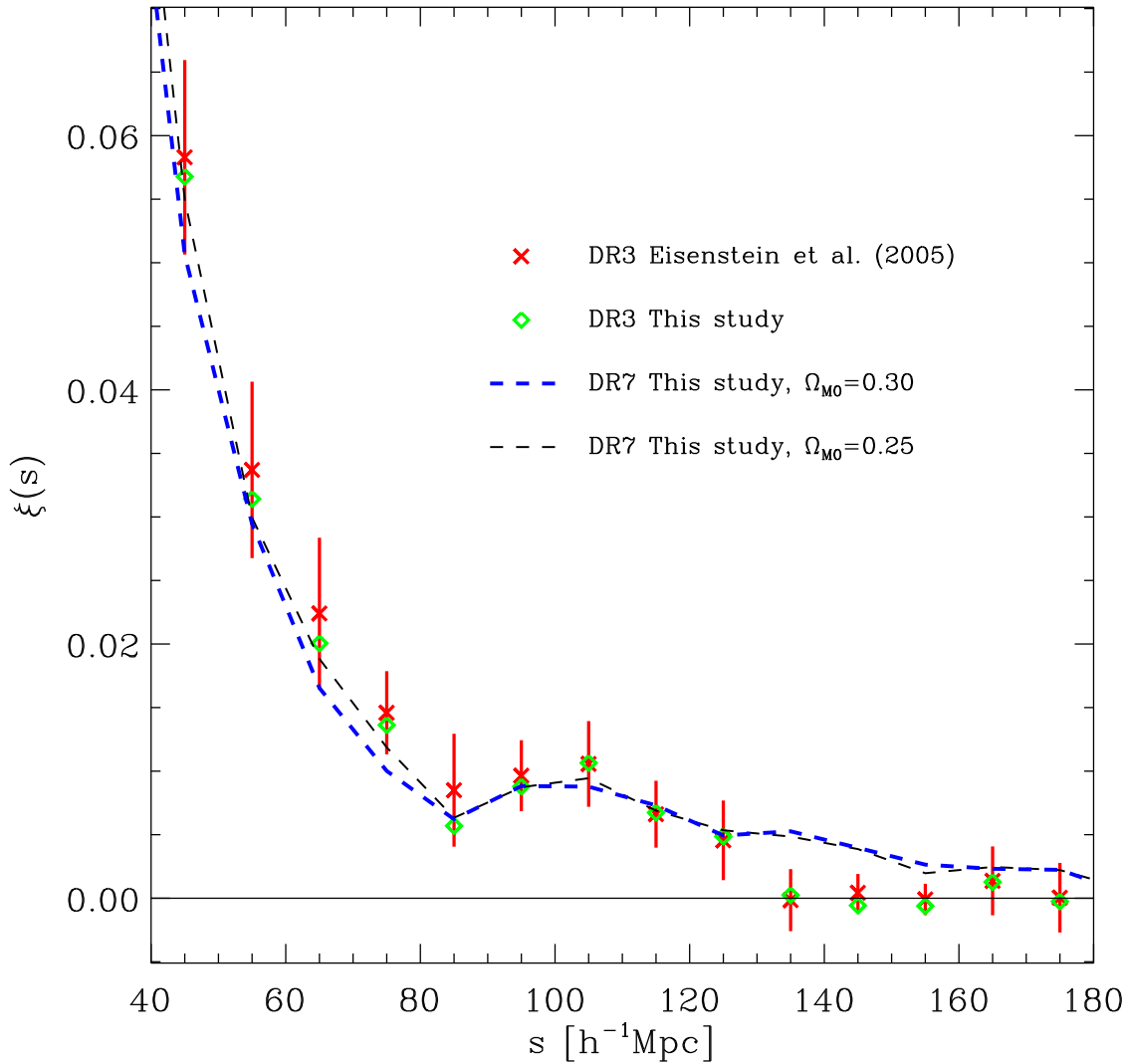}
\caption{DR3 and DR7 $\xi(s)$:  Our DR3 results (green diamonds) show excellent agreement
 on the scales investigated here with those published
by \cite{eisenstein05b} (red crosses and uncertainty bars).
 The remaining discrepancies are consistent with shot-noise in the random
catalogs. The dashed lines are our results for DR7-Full which shows a
stronger clustering signal at $135<s<180$\hmpc. In the thick blue dashed line we used the same
$\Omega_{M0}=0.3$ flat cosmology as the DR3 results, and thin black dashed line 
$\Omega_{M0}=0.25$. Both cosmologies agree very well at scales
discussed here $80<s<200$\hmpcii. In our DR3 result we use number
ratio of random to data points $r \sim 50$ and for DR7 $r \sim 15.6$.}
\label{dr3xis}
\end{center}
\end{figure}

\begin{figure}[htp]
\begin{center}
\includegraphics[width=\textwidth]{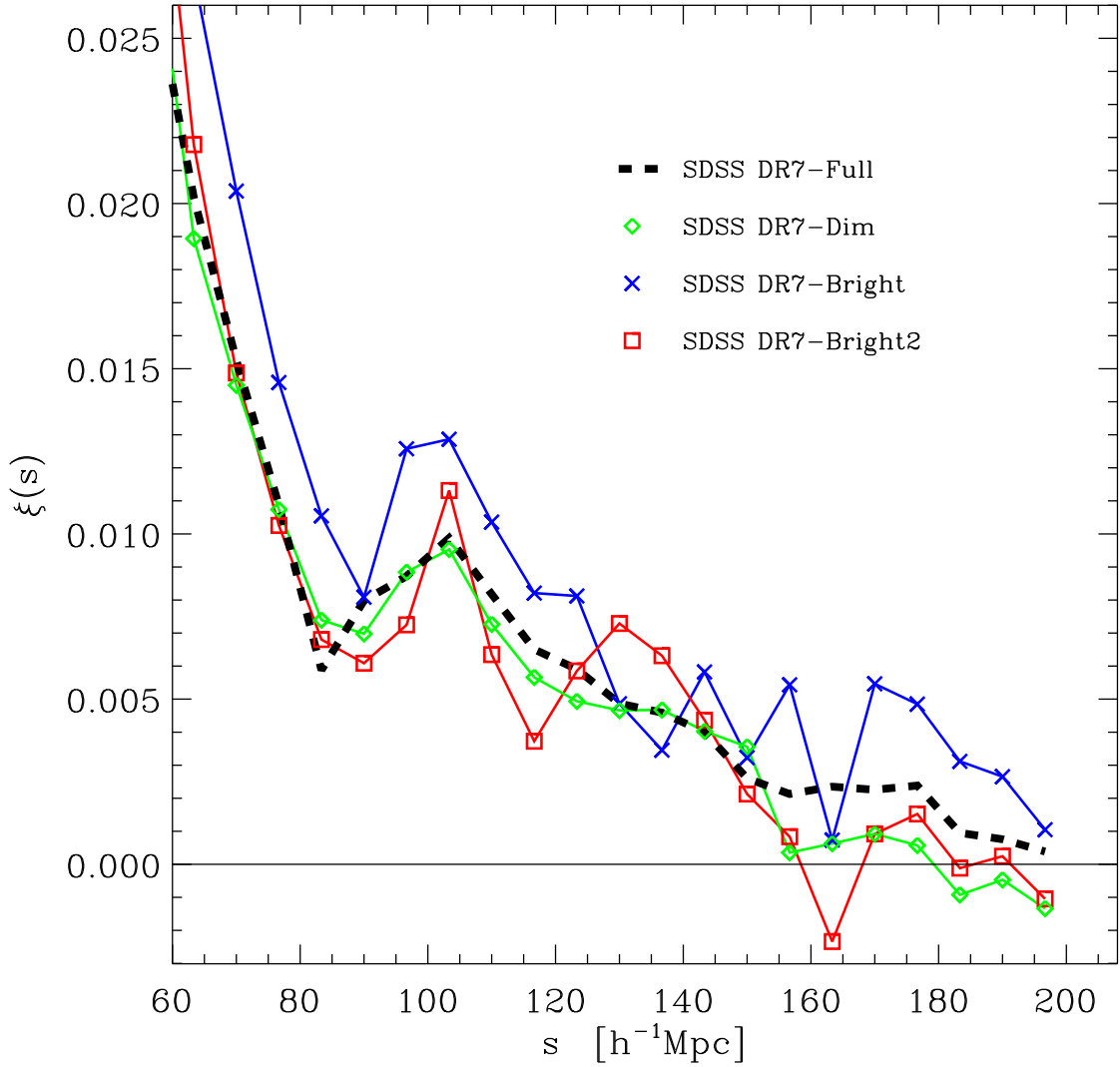}
\caption{DR7 Subsample $\xi(s)$: Comparing DR7-Full (thick dashed line) to
subsamples \qvl (green diamonds), \vl (blue crosses) and \vll (red
squares). \vl shows a stronger signal than the other samples on most
scales. The peak position appears consistent for all
subsamples. For the uncertainties of \qvl and \vl please refer to
Figures \ref{dr7qvl} and \ref{dr7vl}, respectively.}
\label{dr7subsamples}
\end{center}
\end{figure}

\begin{figure}[htp]
\begin{center}
\includegraphics[width=\textwidth]{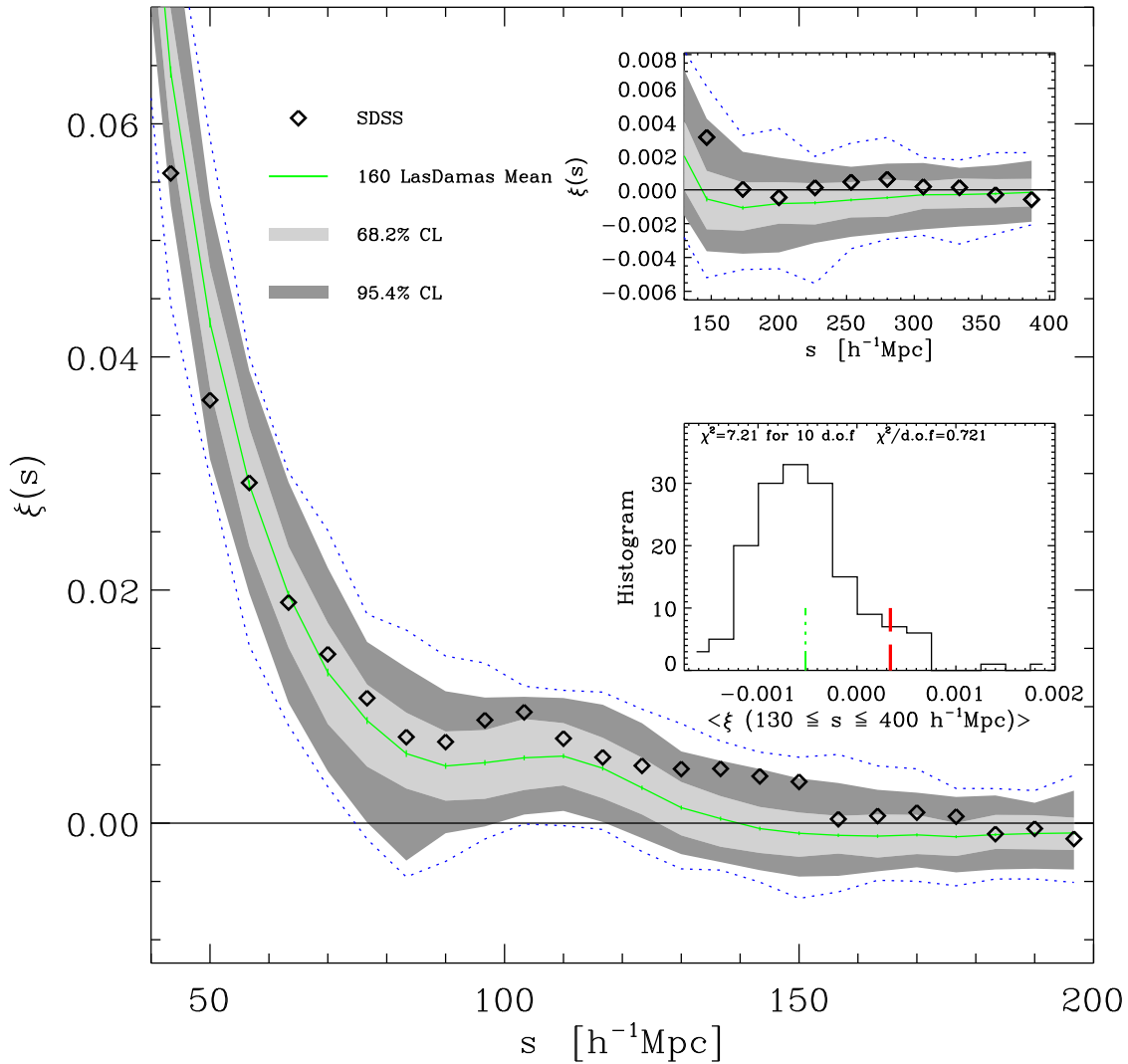}
\caption{\small{\qvl $\xi(s)$: Results from SDSS (black diamonds) and
from the LasDamas mock catalogs. The mock mean $\overline{\xi}_{mock}$
is the green solid line and the uncertainties in the mean are small
vertical green lines. The variance for one realization is presented by
the gray bands: $68.2\%$ light gray ($1\sigma_{mock}$), and $95.4 \%$
by dark ($2\sigma_{mock}$).  The blue dotted lines are the outermost
result of all mocks in each separation bin (not one realization in
particular). Top Inset: Same format as main figure extending the results
with wider bins to larger scales. 
 Bottom Inset: Significance of large-scale
clustering - we average $\xi$ in bins $[130,400]$\hmpc. The observed
$\langle \xi \rangle$ (red thick dashed line) is clearly within $2\sigma$
of the mock realizations (black histogram), with a $\chi^2$ fitting on $10$ d.o.f
yielding $\chi^2/d.o.f=0.72$. The mock mean result is the thin green dot-dashed line.}}
\label{dr7qvl}
\end{center}
\end{figure}

\begin{figure}[htp]
\begin{center}
\includegraphics[width=\textwidth]{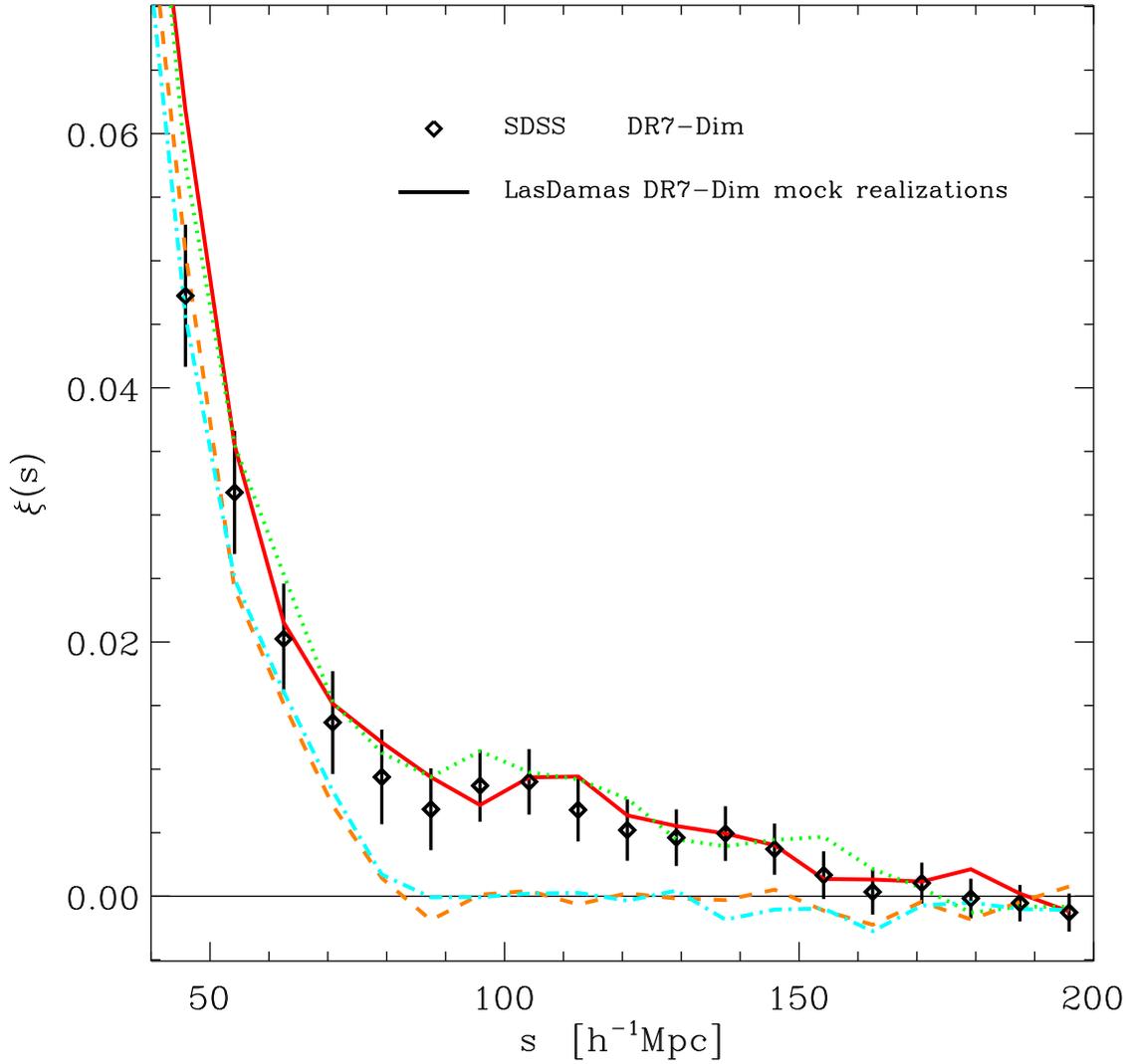}
\caption{\qvl $\xi(s)$ Mock Realizations: Hand-picked LasDamas mocks (lines)
 compared with the SDSS result (diamonds). The green (dotted) and red (solid)
show fairly good agreement on most large-scales with observation,
though their peak position appear to be in different locations.
 The orange (dashed) and cyan (dot-dashed) mocks show example
realizations without a baryonic acoustic feature.
Using a liberal approach, we counted a minimum bound of $\nopeaks\%$ (17) from
the full set of 160 mocks with no sign of a peak.} 
\label{dr7dimmocks}
\end{center}
\end{figure}

\begin{figure}[htp]
\begin{center}
\includegraphics[width=\textwidth]{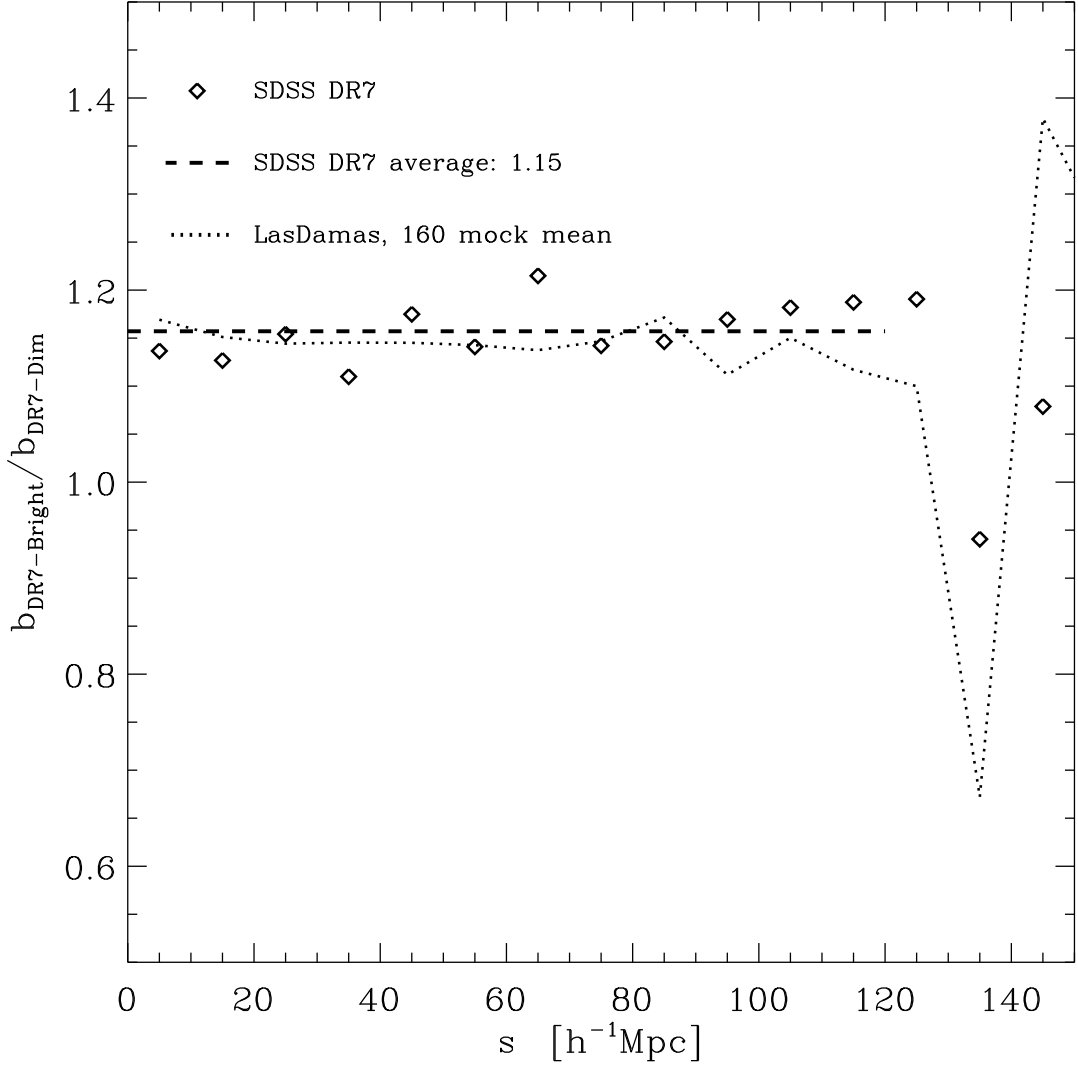}
\caption{Relative Bias:  Comparing redshift-space bias ratio
$b_{DR7-Bright}/b_{DR7-Dim} \equiv \sqrt{\xi_{DR7-Bright} /
\xi_{DR7-Dim}}$ between the two subsamples for both the observed LRGs
(symbols) as well as the LasDamas mock mean (dotted lines). The average 
observed value is taken between $0-120$\hmpc indicated in dashed lines
valued at 1.15.} 
\label{bias}
\end{center}
\end{figure}

\begin{figure}[htp]
\begin{center}
\includegraphics[width=\textwidth]{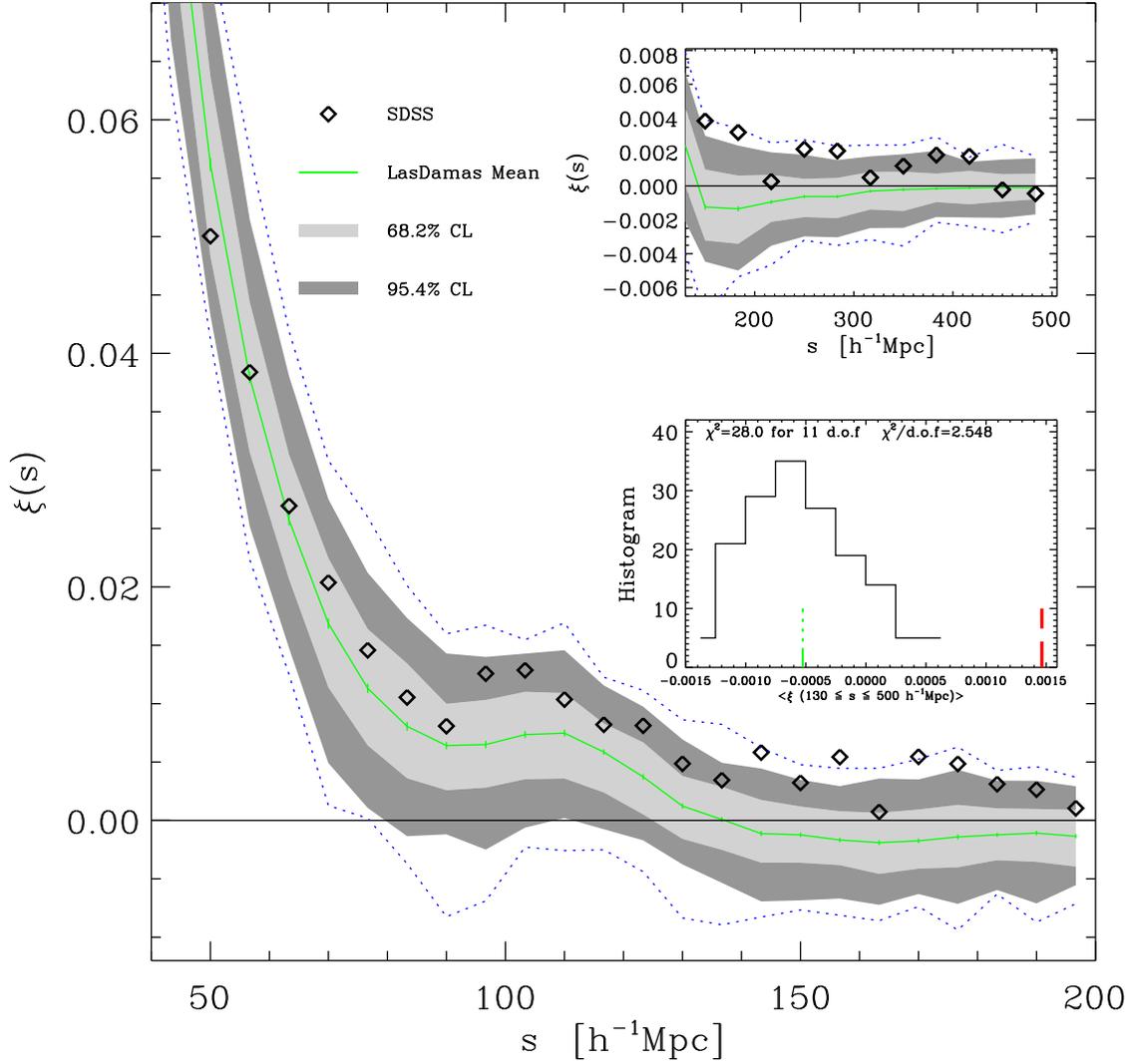}
\caption{\vl $\xi(s)$: Same format as Figure (\ref{dr7qvl}) for the brighter
sample. The insets here extend to $500$\hmpcii. The observed subsample shows a
significantly stronger signal at large scales than produced by our 
$\Lambda$CDM+HOD model. } 
\label{dr7vl}
\end{center}
\end{figure}

\begin{figure}[htp]
\begin{center}
\includegraphics[width=\textwidth]{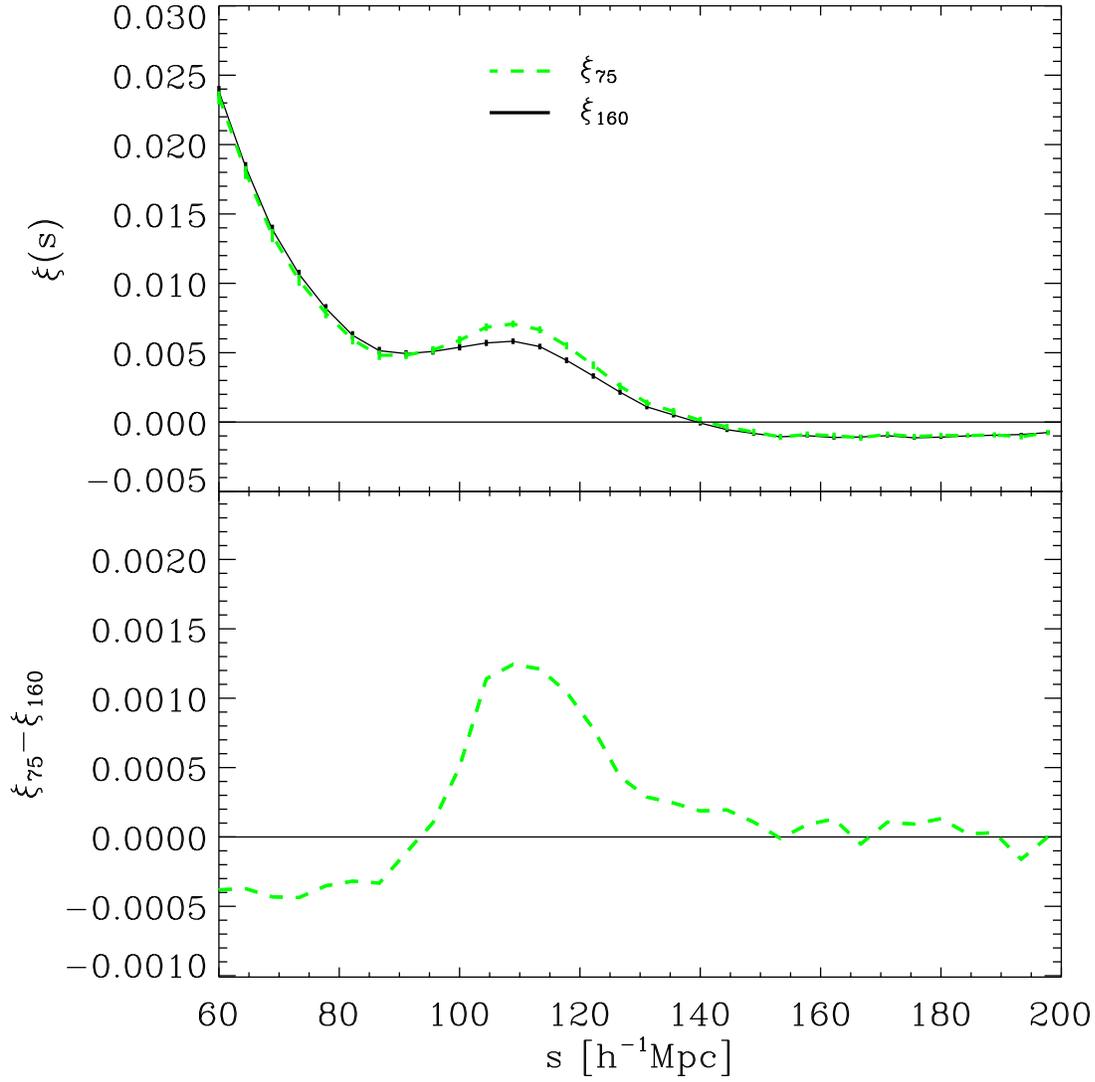}
\caption{Baryonic Acoustic Peak Position in \qvl mocks:
Top Panel- The line for the 75 clear-peaked mocks $\overline{\xi}_{75}$ (bright green dashed) shows a more pridominant peak 
than that of the full catalog $\overline{\xi}_{160}$ (solid black).
Both means show roughly the same peak position, and width.
The vertical lines are $1\sigma$ uncertainties of the mean.
Bottom Panel- Residual  $\overline{\xi}_{75}-\overline{\xi}_{160}$ as function of scale.
}
\label{clearvsall}
\end{center}
\end{figure}

\begin{figure}[htp]
\begin{center}
\includegraphics[width=\textwidth]{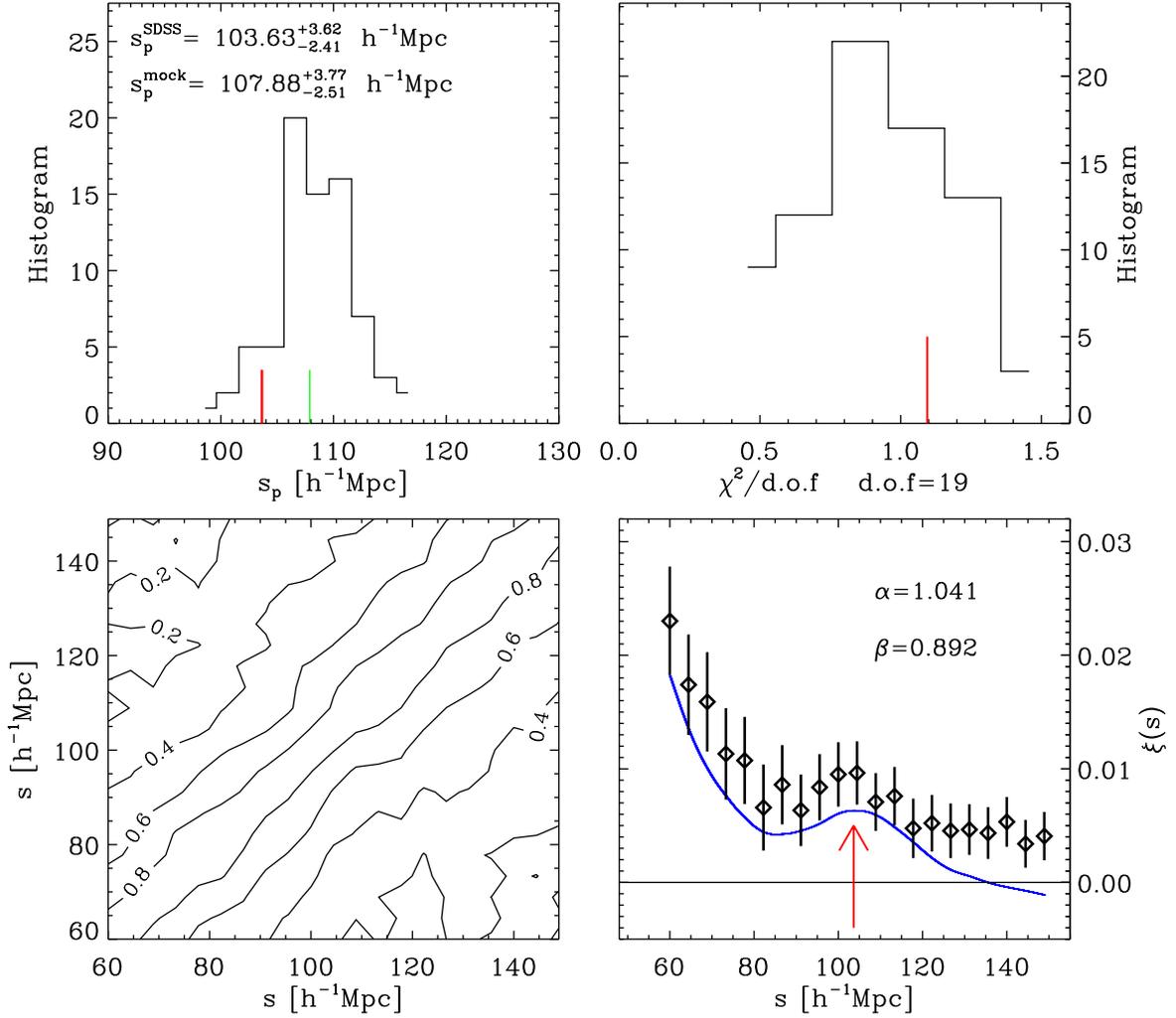}
\caption{Measuring the Baryonic Acoustic Peak Position \speak in \qvl
Sample: Top Left Panel- Histogram of 75 mock realizations with clear sign
of a peak. Vertical red (thick dark) solid line is observed value and green (thin bright) is that of mock mean.
$\sigma_{\pm}=^{3.59}_{2.39}$\hmpc are the $1\sigma$ CL when
counting $68.2\%$ of the mock peaks around the mock mean $107.88$\hmpcii,
normalized to that of the observation by $\sigma \sim
\xi$. 
 Top Right Panel- $\chi^2$ histogram for mock realization fits to
model, where the red line indicates that of the observation at
{\bf{$\chi^2=23.6$}} for $19$ d.o.f. Bottom Right Panel- Fitting the observed
$\xi(s)$ (black diamonds) to model $\beta \overline{\xi}(\alpha \cdot
s)$ (blue line), where $\overline{\xi}$ is the mock mean. Red arrow indicates
\speakii$=$\speakhR\hmpc and is the same as in the top left panel.
 Indicated also are best-fit $\alpha$ and $\beta$
parameters. Black vertical lines are $\sqrt{C_{ii}}$. Bottom Left Panel-
$C_{ij}/(\sigma_i\sigma_j)$. } 
\label{bafposition}
\end{center}
\end{figure}

\begin{figure}[htp]
\begin{center}
\includegraphics[width=\textwidth]{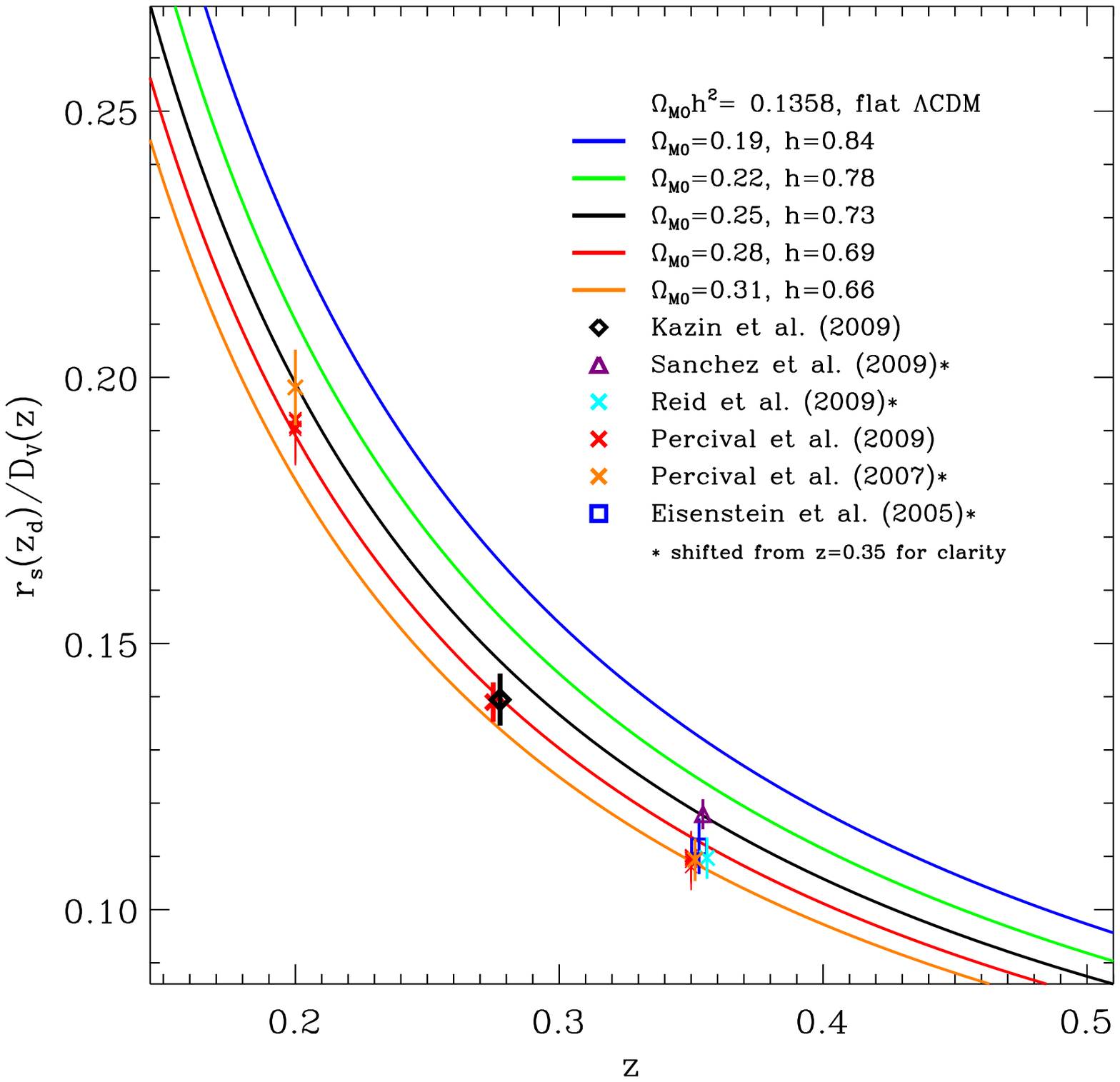}
\caption{$r_s(z_d)/D_V(z)$ Result: We obtain $r_s/D_V(z=0.278)$ of
 \rsDvR$\pm$\drsDvR \ (black diamond; $1\sigma$ uncertainty) in good agreement with the $z=0.275$ result presented by
\cite{percival09a} (red crosses).
Other results (\citealt{sanchez09a}: purple triangles,
\citealt{reid09a}: cyan crosses, \citealt{percival07a}: orange crosses; \citealt{eisenstein05b}: blue square) 
are indicated. These points are not all independent as they use the same sample. The solid lines
show predictions of various flat $\Lambda$CDM cosmologies constraining $\Omega_{M0} h^2=0.1358$
and varying $\Omega_{M0}$ and $h$, where the top (blue) line is $\Omega_{M0}=0.19$, $h=0.84$ 
and the bottom (orange) line is $\Omega_{M0}=0.33$, $h=0.64$. Intermediate steps shown are 
$\Omega_{M0}=0.22,0.25,0.29$.  Our result clearly agrees with ranges $\Omega_{M0}=[0.25,0.33]$
 and $h=[0.64,0.73]$.}
\label{Dvmeasure}
\end{center}
\end{figure}

\begin{figure}[htp]
\begin{center}
\includegraphics[width=\textwidth]{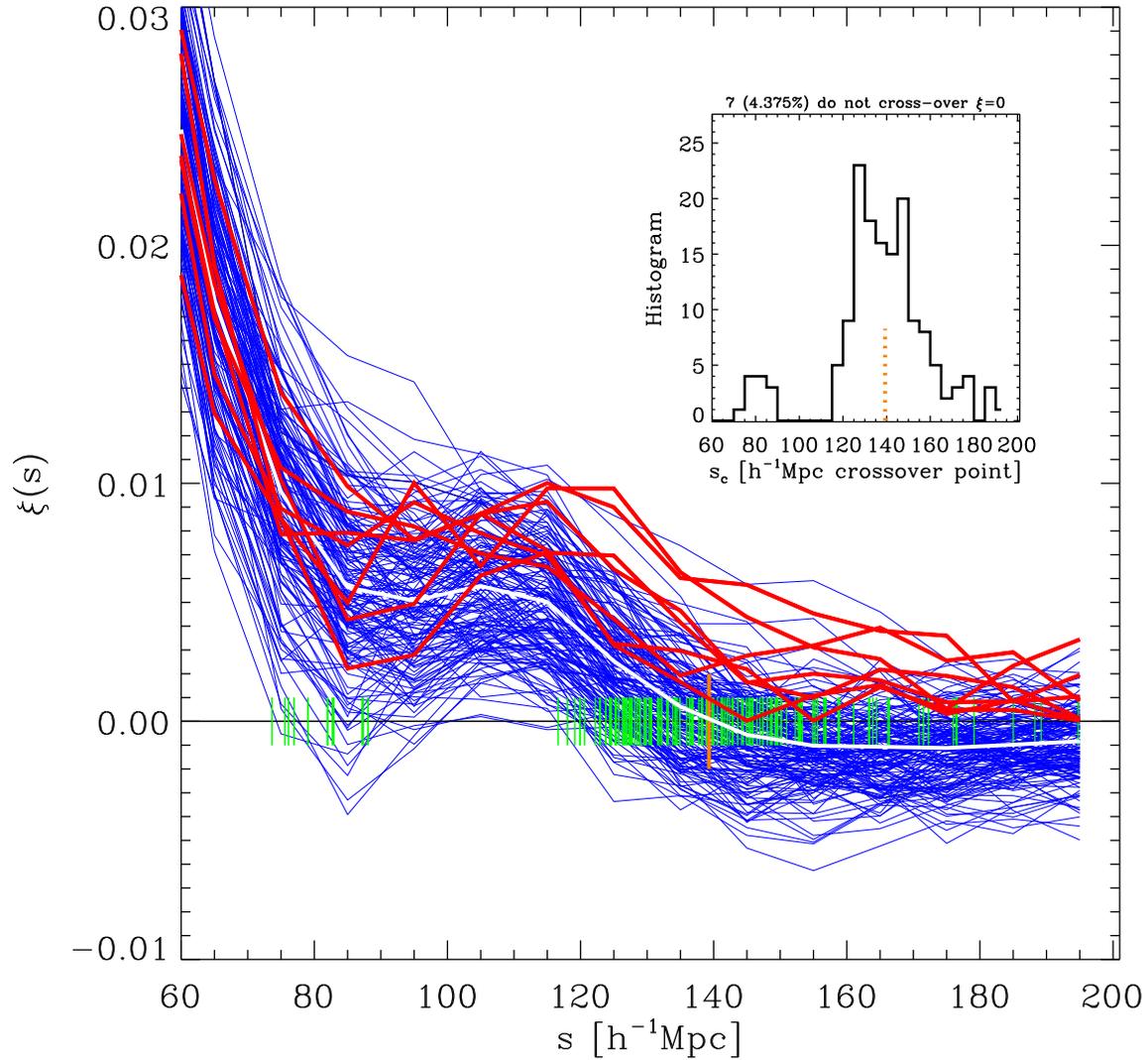}
\caption{$\xi(s_c)=0$: Here we show all 160 \qvl mock realizations.
 Those with a crossover point $s_c$ (crossover marked in vertical green lines) are blue solid lines, 
and the $4\%$ without are in thick red. The mean value is the solid white line, and its $s_c\sim 140$\hmpc
is indicated by the vertical orange line. In the inset we show a histogram of all $s_c$  values, where the dotted orange line is the mean value.}
\label{crossover}
\end{center}
\end{figure}

\begin{figure}[htp]
\begin{center}
\includegraphics[width=\textwidth]{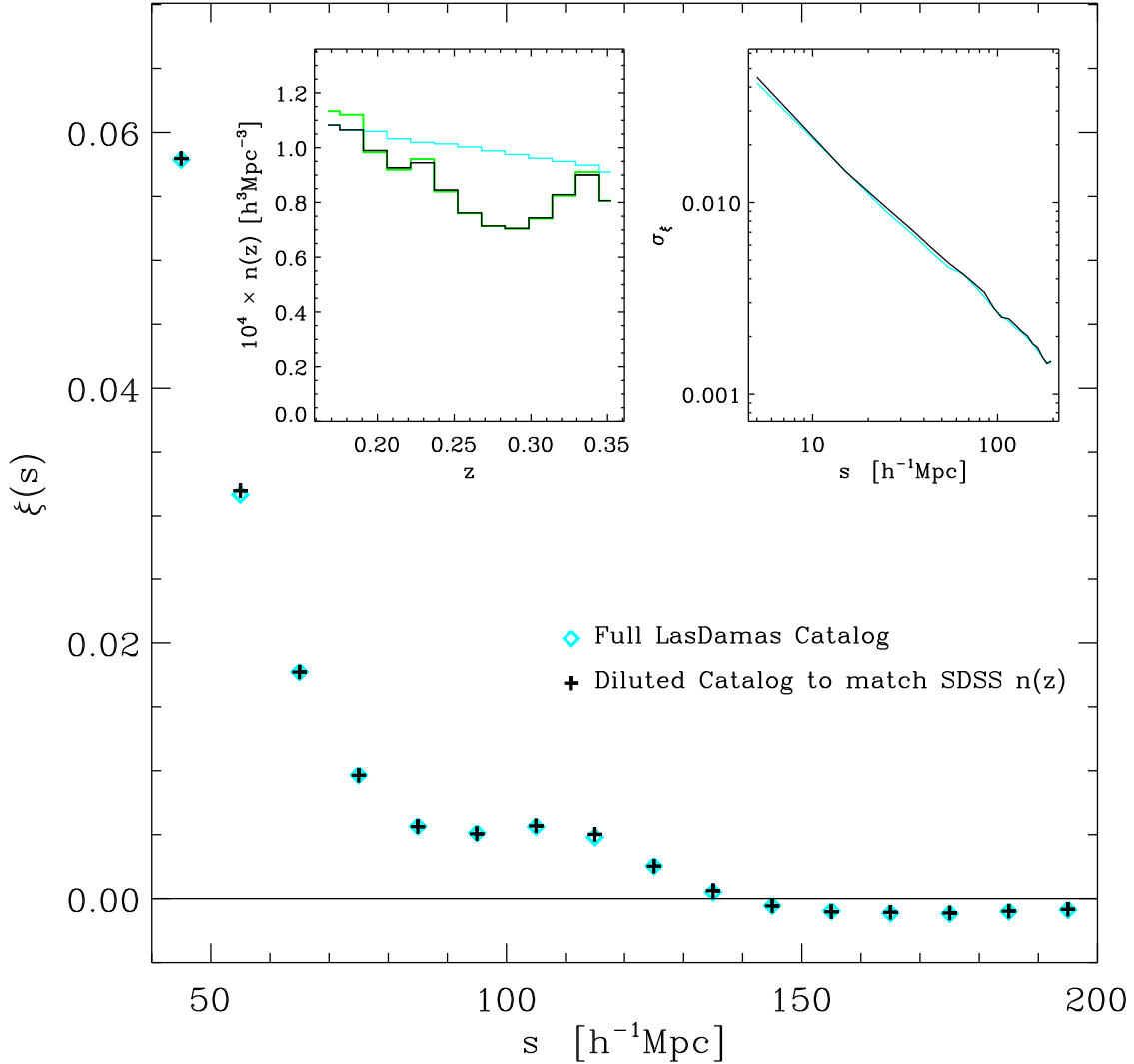}
\caption{$n(z)$ effect on $\xi$ and $\sigma_\xi$ in \qvlii: 
Left inset shows $n(z)$ of LasDamas mock catalogs (light cyan histogram; $n(z)_1$)
which has a slight negative slope, and that observed in the SDSS sample 
(thick green histogram, same as in Figure \ref{nofz}; $n(z)_\mathrm{SDSS}$). We exclude $15\%$ of the mock
 galaxies in each realizatoin to match $n(z)_\mathrm{SDSS}$. The result is the 
black histogram ($n(z)_2$). In the main plot the mean $\xi$ over 160 mock \qvl realizations 
for the different selection functions. Cyan diamonds are results of $n(z)_1$ and the black crosses are
for $n(z)_2$. Both $\xi(s)$ are radially weighted with their respective $n(z)$, and are in very good
agreement. Right inset shows the same for the r.m.s $\sigma_\xi$, indicating a
slightly stronger variance for the latter sample, as it has slightly
higher Poisson shot-noise.} 
\label{xisfeffectplot}
\end{center}
\end{figure}

\begin{figure}[htp]
\begin{center}
\includegraphics[width=\textwidth]{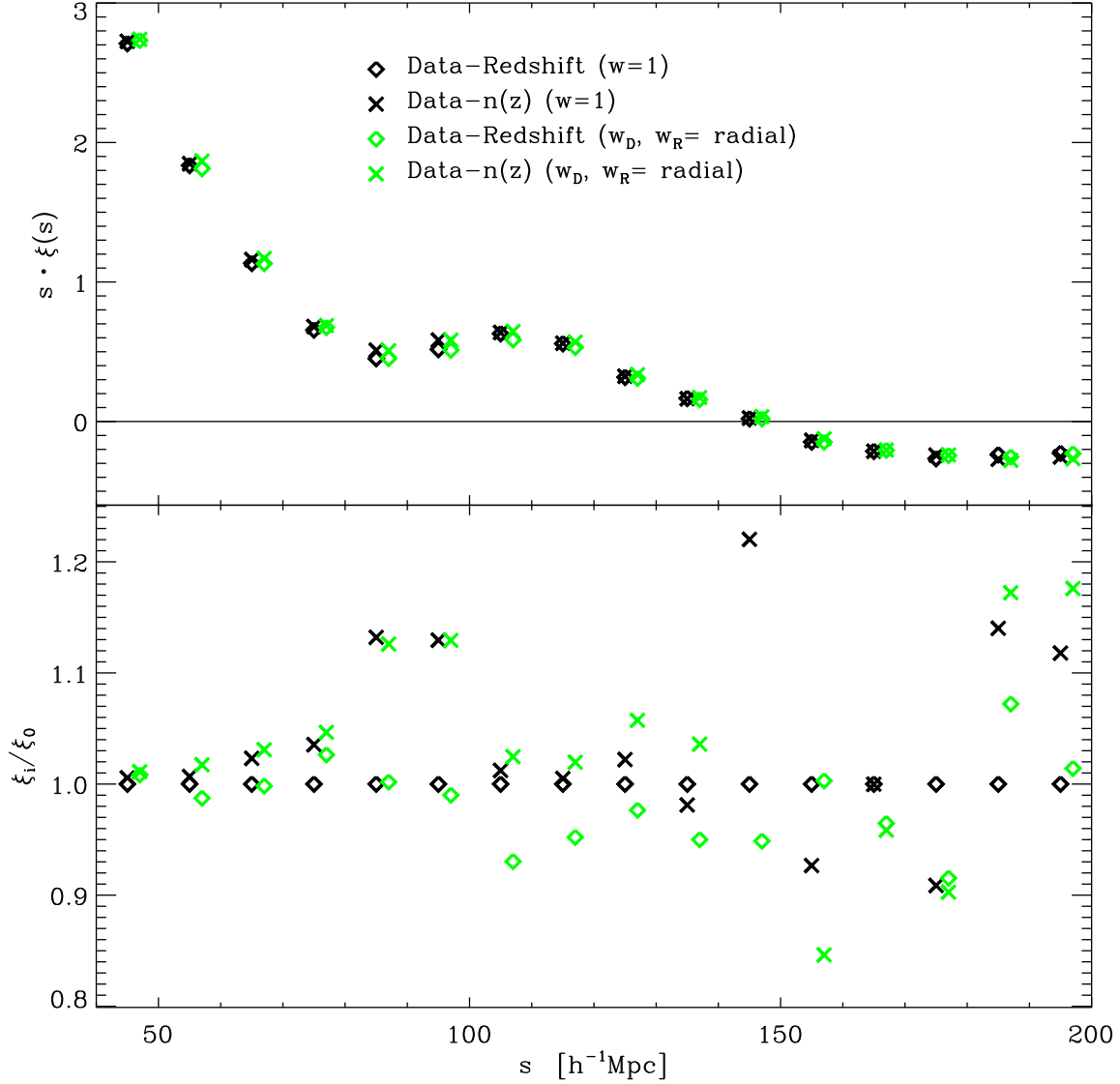}
\caption{Random point distribution effects on $\xi$ in \qvlii: Results
for average over 8 mock realizations when using original mock $n(z)_1$
(Figure \ref{xisfeffectplot}). 
In each realization we  
distribute radial distances to random points differently:
 diamonds- data point redshifts distributed randomly to
random catalog (Data-Redshift),
 crosses- random $n(z)$ matches data but otherwise independent redshift
distribution (Data-$n(z)$).
Radial weighting schemes: 
Black- no weighting , Blue- both data
and random are weighted radially in same manner (shifted to right by $2$\hmpc for clarity),
Top Panel- To clarify differences we plot $s\cdot \xi(s)$.
Bottom Panel- Ratio of each case $\xi_i$ to unweighted Data-Redshift ($\xi_0$).
We clearly see that Data-Redshift (diamonds) yield the lowest result, 
hence diminishing the radial clustering mode. This is noticeable in range
 to $50<s<130$\hmpc after which results agree with
most of the other options. Weighting causes no effect in this
case as the $n(z)$ used is close to volume-limited.
}
\label{randdistribute}
\end{center}
\end{figure}

\begin{figure}[htp]
\begin{center}
\includegraphics[width=\textwidth]{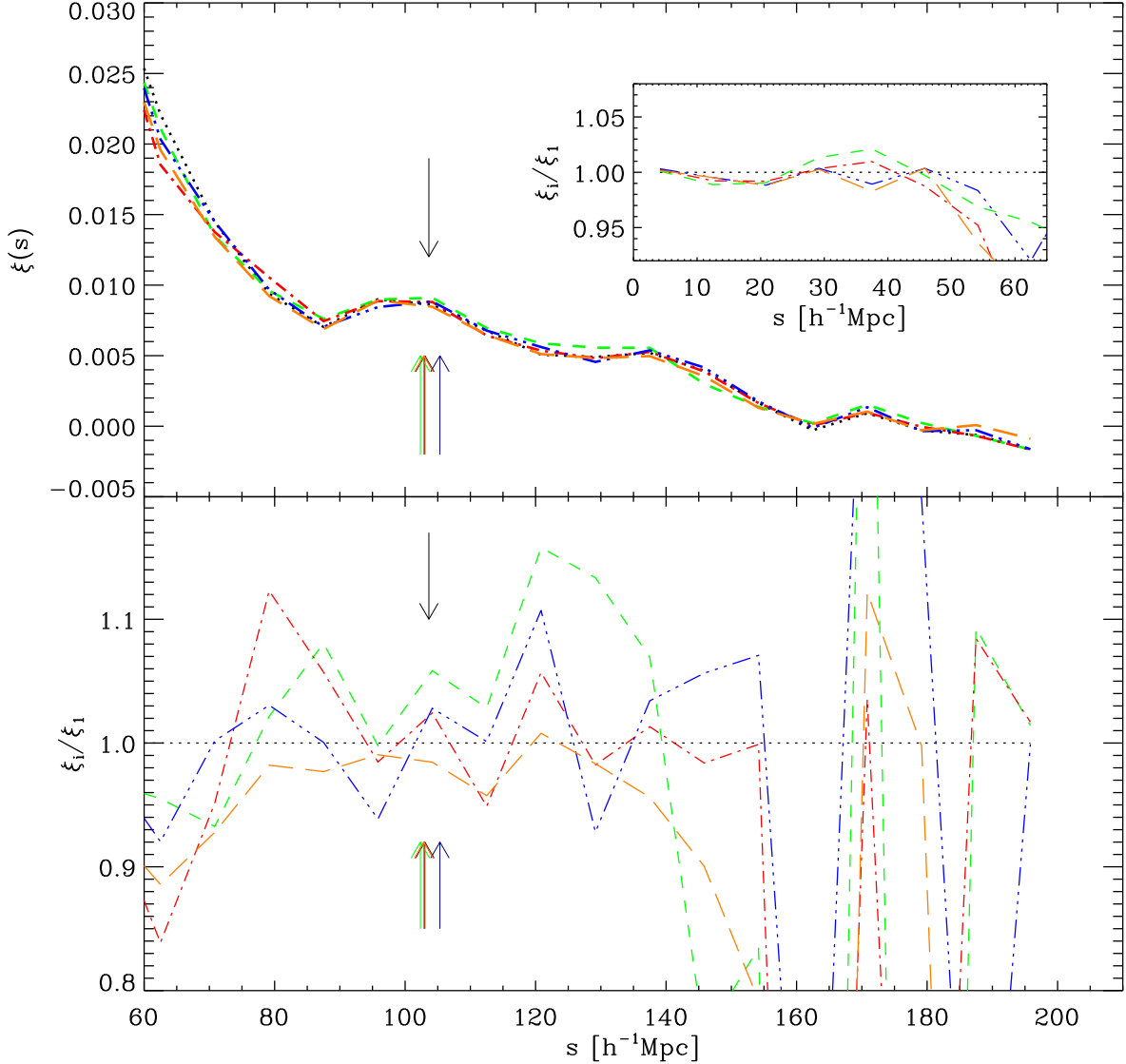}
\caption{Random Shot-Noise: We run the same algorithm 
on the observed \qvlii, using five different random catalogs,
each in turn.
The $\xi$ results are given in the top panel, where the five random sets
are represented by: black-dotted, green-dashed, red-dot-short-dashed,
orange-long-dashed, blue-triple-dotted-dashed lines. To facilitate differences
the bottom inset shows the ratio of 
each $i^{\mathrm{th}}$ catalog $\xi_i$ in respect to the first random catalog $xi_1$.
The inset has the same format as the bottom panel for smaller sacles.
The bottom arrows in both panels pin-point the peak position found,
according to our algorithm. We see that four random catalogs
result in same \speak within $1$\hmpc where a fifth is $3$\hmpc
larger than the smallest obtained value.
These results are for a ratio of random to data points of $r=$N$_R/$N$_D\sim 15.6$.
To reduce this effect in our main analysis we use $r\sim 50$
(its result is marked by top arrow at \speak$=\speakhR$\hmpcii). 
} 
\label{diffrandcat}
\end{center}
\end{figure}

\begin{figure}[htp]
\begin{center}
\includegraphics[width=\textwidth]{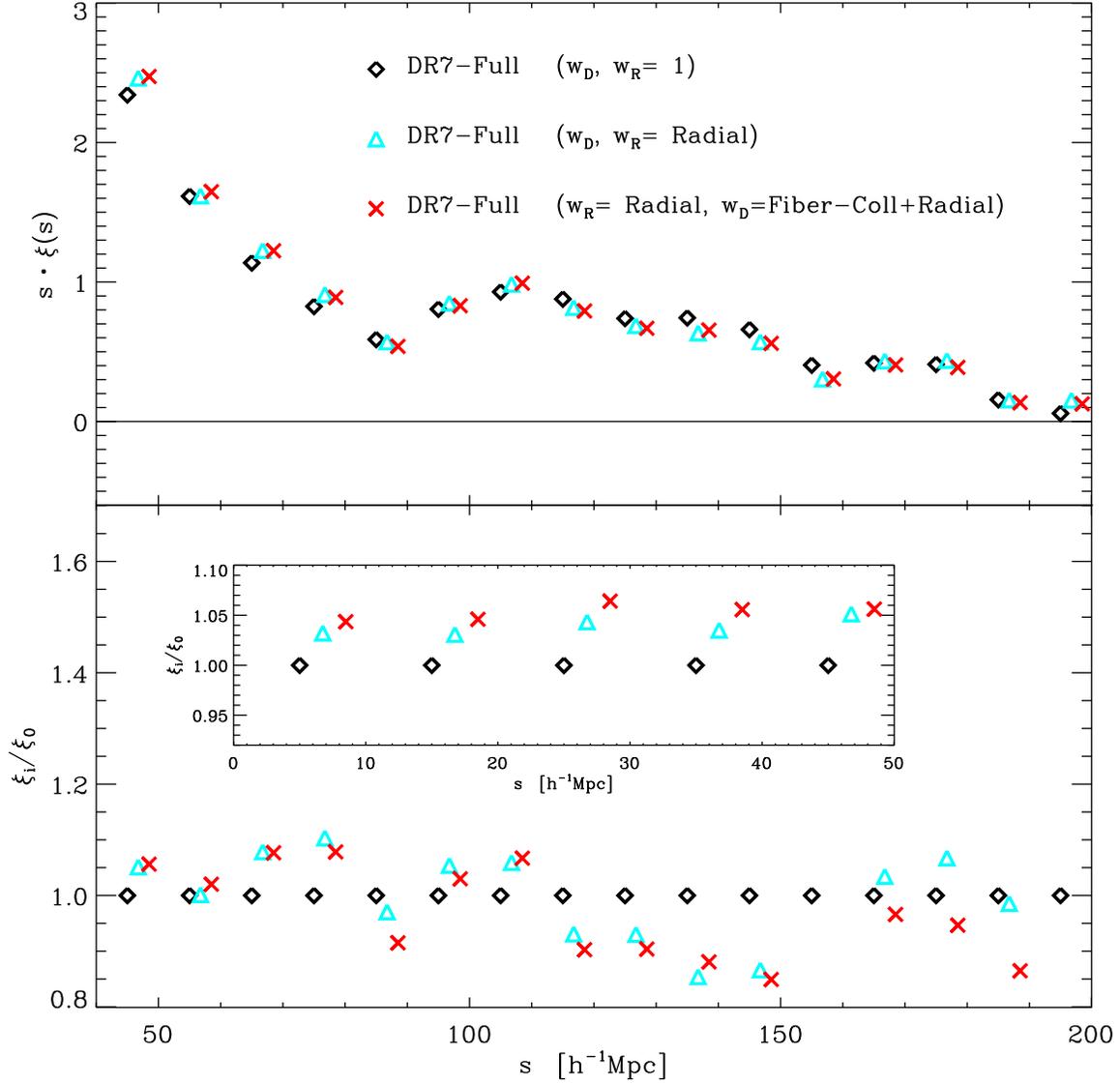}
\caption{Weighting Effects on $\xi$: We compare different weighting
schemes on the observed DR7-Full. 
 $w_D$ and $w_R$ indicate the weight used on data and
random points respectively, where black (diamonds) indicates no weight used ($w=1)$,
cyan (triangles) indicates radial weight (Radial; shifted by $1.75$\hmpc for clarity) and red (crosses) 
indicates radial+(fiber-collision
correction) (Fiber-Collision+Radial; shifted by $35$\hmpc). All options take into account sector completeness corrections.
For clarity we plot in the top panel $s\cdot\xi(s)$ and on the bottom the ratio
of each weighting option $\xi_i$ over the first ($w_D=w_R=1$; $\xi_0$).
}

\label{weightplot}
\end{center}
\end{figure}

In order to calculate the separation in comoving space between galaxy
pairs, we must use the observed redshifts and assume a certain
fiducial cosmology.  For most of this study we assumed a flat
$\Lambda$CDM model with $\Omega_{M0}=0.25$. When analyzing differences
between DR3 and DR7-Full we also tested the cosmology assumed by
\Epaper $\Omega_{M0}=0.3$. We also examine each cosmology on each
subsample and found that the cosmology does not affect the resulting
$\xi$ or the measured \speak significantly relative to our other
uncertainties.  For a direct comparison of both cosmologies in DR7 see
Figure \ref{dr3xis}.

We also tested how the choice of cosmology affects the selection of
LRGs, due to the fact that $M_g$ is sensitive to cosmology.  Within
\qvl ($0.16<z<0.36$, $-23.2<\mathrm{M_g}<-21.2$) we count $61,899$
LRGs when using $\Omega_{M0}=0.25$ and $61,102$ when
$\Omega_{M0}=0.30$, a $1.3\%$ difference. When probing the full
sample, DR7-Full, we find an agreement in number of LRG selected in
the two cosmologies to better than $1\%$.  This implies that for large
range of redshift an approximate cosmology should not change the
number count very much.

\begin{figure}[htp]
\begin{center}
\includegraphics[width=\textwidth]{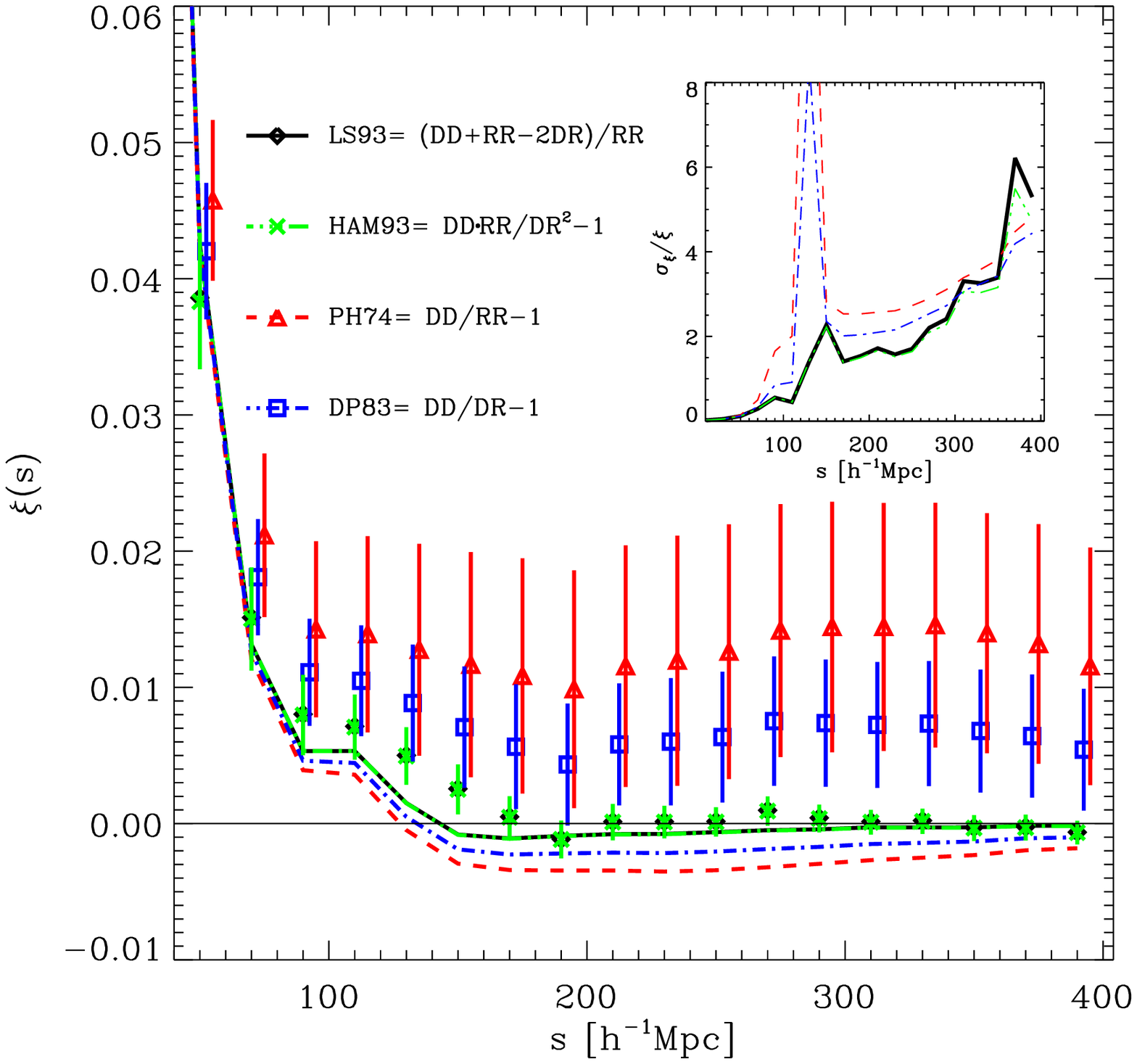}
\caption{$\xi$  Estimators at Large Scales of \qvlii: Comparing 
estimators proposed by LS93 (\citealt{landy93a}; black), HAM93
(\citealt{hamilton93}; green), DP83 (\citealt{davis83}; blue, symbols shifted for clarity)
 and PH74 (\citealt{peebles74}; red, symbols shifted).
 The lines are mock mean values, and
the symbols are those observed. The uncertainty bars on the symbols
are the r.m.s of the mocks. Although the observed results appear in
reverse order than the mock mean, we verified by looking at particular
mock realizations that some are in this order as well.
 Inset: shows the Noise to Signal
$\sigma_\xi/\xi$. 
 The strong spikes are due to sensitivity at the
signal zero crossing. Refer to Figure \ref{rmsplot} for
$\sigma_{\xi}(s)$ of the different estimators.
We find that LS93 and HAM93 agree very well
on large scales, and asymptote quicker to zero
than the other two. The ratio of random to data
points used for SDSS is $r\sim 15.6$ and for mocks
$r\sim 10$.} 
\label{estimatorplot}
\end{center}
\end{figure}

\begin{figure}[htp]
\begin{center}
\includegraphics[width=\textwidth]{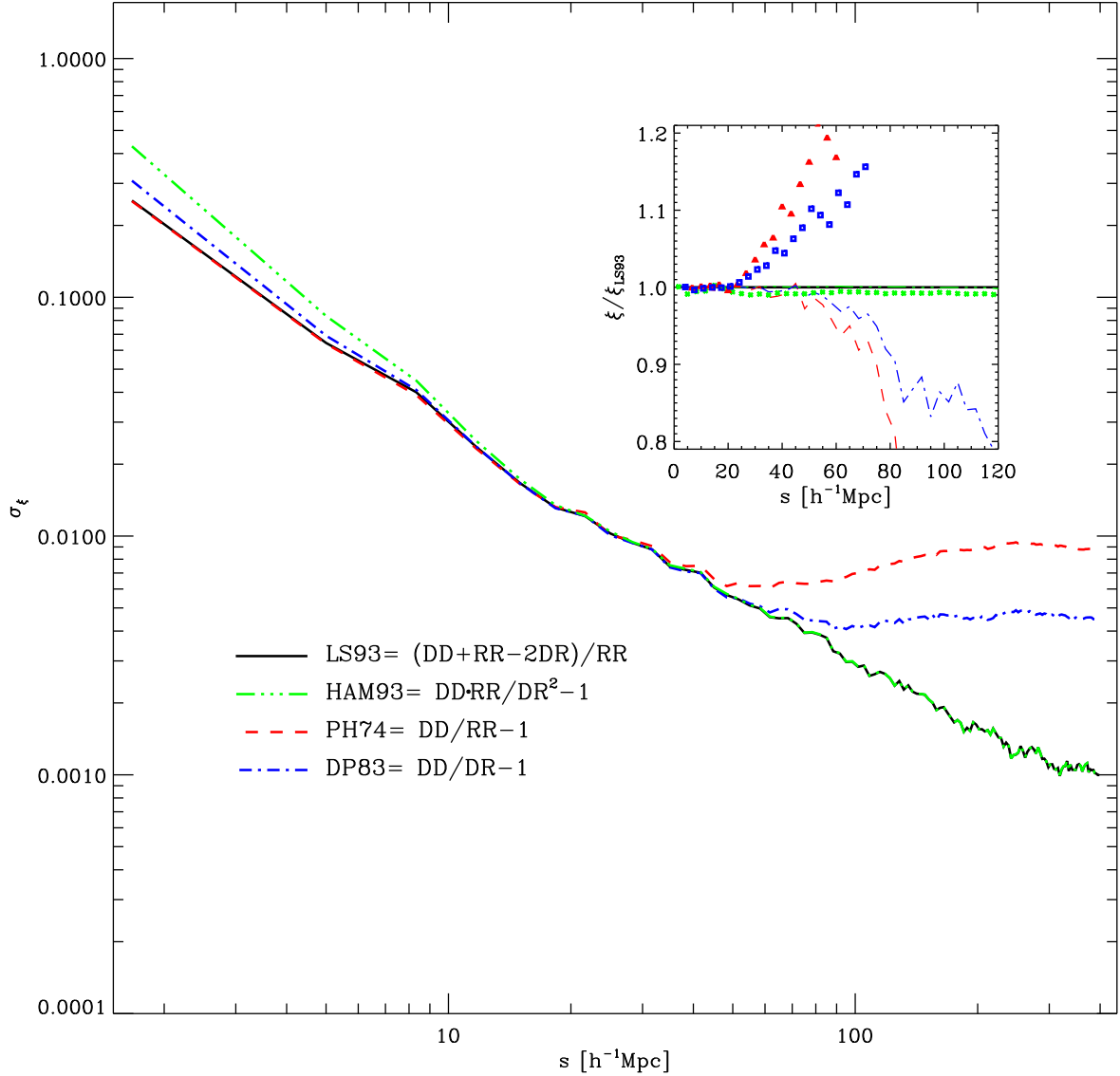}
\caption{Estimator Variance: Main Plot: Same as Fig. \ref{estimatorplot}
showing the r.m.s $\sigma_{\xi}=\sqrt{C_{ii}}$ where $C_{ij}$ is the
covariance of $\xi$. PH74 and DP83 show poor variance
performance on large-scales. HAM93 agrees very well with 
LS93 on large scales but has a larger variation on $\lesssim 10$\hmpcii.
Inset: Comparing estimators to \cite{landy93a}.
The ratio of random to data 
points used for SDSS is $r\sim 15.6$ and for mocks 
$r\sim 10$.
}
 
\label{rmsplot}
\end{center}
\end{figure}

As mentioned in \S \ref{method}, the $\xi$ calculation requires the
choice of a particular estimator. We tested the \cite{landy93a}
estimator against those proposed by \cite{hamilton93}, \cite{davis83},
\cite{peebles74}. Figures \ref{estimatorplot} and \ref{rmsplot} 
show our $\xi$ and $\sigma_{\xi}$ results, when using \qvlii. In Figure
\ref{estimatorplot} the mock mean (over 160 realizations) are the lines,
 while the SDSS result are the symbols. The different colors indicate the different
estimators used, where \cite{landy93a} (black solid, LS93 hereon) and
\cite{hamilton93}(green triple-dot-dashed, HAM93)
are not distinguishable by eye for the most part. The inset shows the
noise-to-signal ratio.  Bear in mind that the spikes around $140$\hmpc
simply result from the signal crossing zero around that scale.  
In Figure \ref{rmsplot} the notation is the same when
plotting $\sigma_{\xi}(s)$ for the different estimators. The inset in
this plot shows the ratio of each estimator relative to that of LS93:

We find the following: 
\begin{enumerate}

\item{Mean value $\overline{\xi}$: }  Averaging over all 160 mocks we
see that LS93 and HAM93 agree on all scales. The inset of Figure \ref{rmsplot}
shows that \cite{peebles74} (PH74) deviates from the latter by $10\%$ at 
scales of $\sim 75$\hmpc and \cite{davis83} (DP83) does the same at $85$\hmpcii.
We also note that LS93 and HAM93 asymptote faster to zero than the others.

\item{r.m.s $\sigma_{\xi}$:  } The r.m.s of the signal varies among
the estimators on various scales. PH74 yields the same $\sigma_\xi$
as LS93 up to scales of $\sim 40$\hmpc before it starts strongly branching off.
This is clearly seen in the main plot of Figure \ref{rmsplot}, and 
indicated in the other plots by the strong variation of the observed 
result. Both HAM93 and DP83 have a larger variance than LS93 at scales
smaller than $10$\hmpcii. HAM93 later matches the LS93 on all larger 
scales very well, where DP83 breaks off at $\sim 60$\hmpcii.

\end{enumerate}

We conclude that the \cite{landy93a} and \cite{hamilton93} estimators
agree in signal for each realization on large scales, and perform much
better than the other two as their variance is smaller, and converge
much more quickly to zero. In particular, the other two estimators
yield very large uncertainties on large scales for individual
realizations. We also find that \cite{hamilton93} does not perform as
well as \cite{landy93a} on smaller scales $s<10$\hmpcii.  Our analysis
on \qvl agrees with the other samples \vlii,
\vlii-Near, and \vlii-Far. 
Here we use ratio of random points to data of $r \sim 15.6$
for the SDSS and $r \sim 10$ for mocks. 
\cite{kerscher00a} shows \cite{landy93a} to be the estimator with
the best preformance relative to the true 
$\xi$ in periodic box measurements.

To test the importance of binning, we use the
\cite{landy93a} on all 160 mock \qvl realizations and find a strong dependence
of the variance on the bin size, as expected. We test various bin
widths $\Delta s$ between $\Delta s=[0.55,10]$\hmpc noting that
variance changes strongly, where the lowest $\Delta s$ yields the
noisiest $\sigma_\xi$, due to the higher shot-noise in each bin.  For
most of our analysis, including the \bafii, we used $\Delta
s=4.44$\hmpcii. We find that peak position \speak is not altered by
more than $1\sigma$ for any choice of bin size.

\begin{figure}[htp]
\begin{center}
\includegraphics[width=\textwidth]{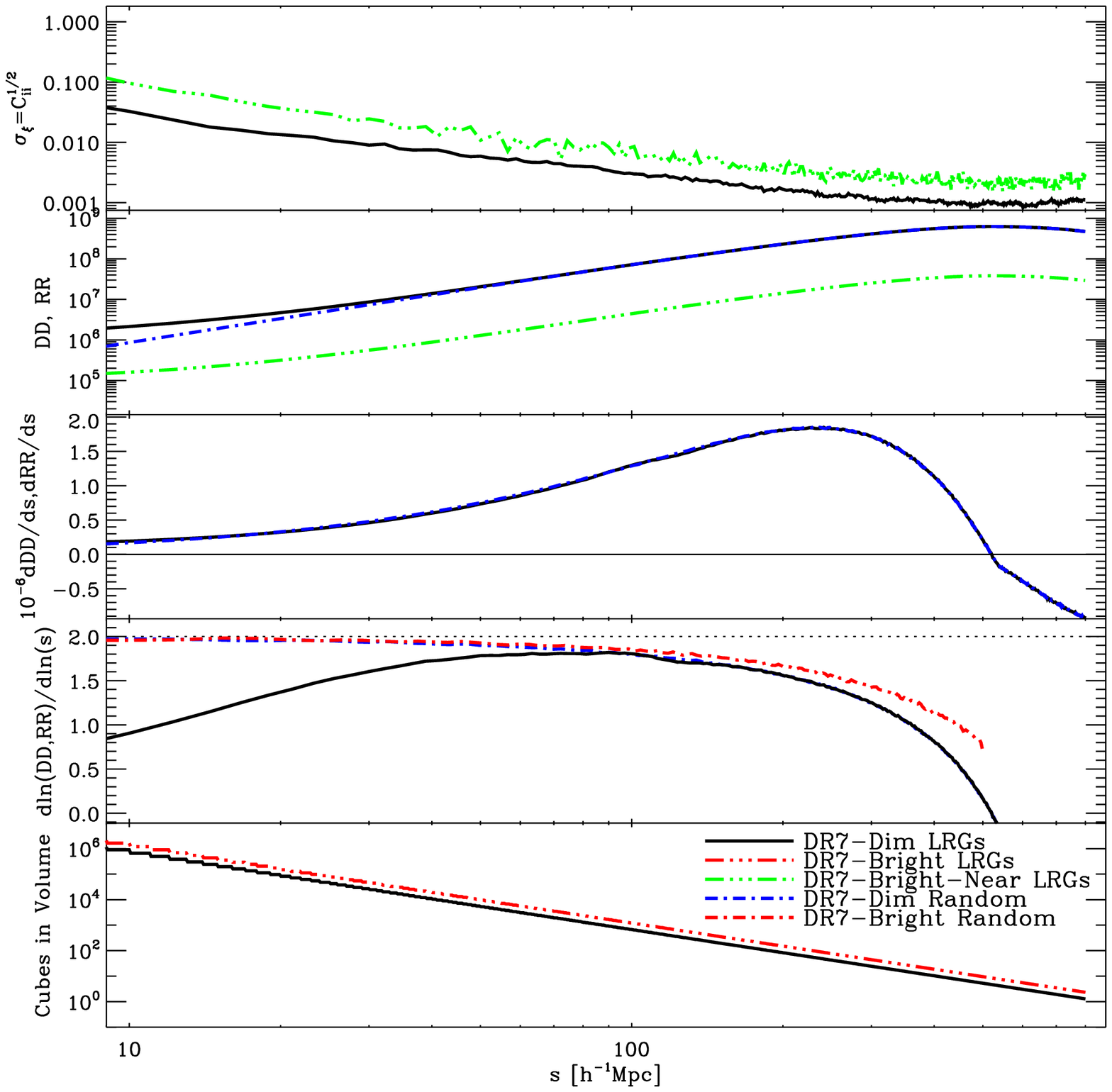}
\caption{LRG and random point statistics up to $s=800$\hmpcii. 
a) signal r.m.s $\sigma_\xi$ for \qvl (black solid line) and \vlii-Near (green triple-dot-dashed).
b) Normalized data-data pairs DD for \qvl and \vlii-Near (same notation) and random-random of \qvl (blue dot-dashed)
c) Same as (b) only derivative $dx/ds$. where $x=$DD,RR.
d) Examining boundary effects in pair counting. A periodic box would yield $dln(RR)/dln(s)=2$. Hence
deviations from $2$ (dotted line) are effects of boundary.
e) Counting the number of cubes of length $s$ fit in the survey volume, for \qvl (solid black) \vl (red triple-dotted-dashed). 
} 
\label{longscales}
\end{center}
\end{figure}

We also check for boundary limits in our survey, to get an idea of at
what point our $\xi$ estimates are encountering the edges of the
survey. We show in Figure \ref{rmsplot} a negative slope for
$\sigma_\xi$ all the way to $s<400$\hmpcii. A region we definitely
want to avoid in analysis is one in which $\sigma_\xi$ has a positive
slope, which indicates an upper limit to the effective scaling given
the survey volume.  For this reason in the top panel of Figure
\ref{longscales} we continue this plot for \qvl (black solid) and
\vlii-Near (green triple-dotted-dashed) to $800$\hmpc. These 
subsamples (both limited to the range $0.16<z<0.36$) have a declining
$\sigma_\xi$ to $500$\hmpcii, and a slightly positive slope
thereafter.  In the second plot from the top we show the normalized
data-data pairs $DD$ of \qvl and \vlii-Near (same notation), as well
as the random-random pairs $RR$ count of \qvl (blue dot-dashed).  The
dip feature at $\sim 500$\hmpc mentioned before appears here as the
peak at that scale. In the third plot from top we differentiate the
previous results by $s$ to better see where the number of pairs stop
growing with radius.  We see a clear crossover between
$500-600$\hmpcii.

The bottom two plots of Figure \ref{longscales} can be considered
``sanity checks'', as we verify basic statistics. Given a periodic
box we expect the number of random points $N$ around a given a point
at radius $s$ in a spherical shell of width $ds$ to be: $N=4\pi s^2 ds
\overline{n}$ where $\overline{n}$ is the mean density.  $RR(s)$, is
expected to go as $\sim N(s)\cdot N_{R}/2$, where $N_{R}$ is the
number of random points, and $2$ takes into account double
counting. Hence $d\ln(RR)/d\ln(s)$ should yield $2$. Deviations from
this result are due to boundary effects.  In the fourth from top plot
we present $d\ln(x)/d\ln(s)$, where $x$ is $RR$ (blue dot-dashed) and
$DD$ (solid black). The important result from this plot is the
deviation of $d\ln(RR)/d\ln(s)$ from $2$ at large-scales. A $5\%$
difference is noticed at $50$\hmpcii. At the \baf scale the deviation
from $2$ is $\sim 10\%$, at $200$\hmpc by $\sim 23\%$ and at
$400$\hmpc by $60\%$. 
Although this panel
indicates our volume is far from ideal for analysis of $400$\hmpcii,
the LS93 estimator appears to be valid as it corrects for these
boundary effects by comparing number of data pairs to number of
{\it{expected}} random points given boundary conditions.
We preformed the same test for random points within 
a volume of \vl (red dot-dashed) where we extend 
measurement to $500$\hmpcii.

In the bottom plot we determine how many cubes of length $s$ would fit
into the volume contained within \qvl (black solid) and \vl (red
three-dot-dashed). Our choice of $s=400$\hmpc yields $10.31$ for the
former and $18.65$ for the latter. For \vl up to $s=500$\hmpc we count
$9.55$ cubes.  As for the peak position at $s=103$\hmpc, we count
$604$ cubes for the \qvl volume and $1093$ for a \vl volume.

All the plots in Figure \ref{longscales} indicate that our choice of
$s<400$\hmpc is a fair one within the given sample, and does not
depend too much on edge effects.

We also test for uncertainties $\sigma_\xi$ of the observed \qvl and \vl against those
predicted by Gaussianity plus shot noise (\citealt{bernstein94}) 
 at large scales,
 
\begin{equation}
\sigma^2_\xi(s)={2\over (2\pi)^3V} \int d^3k\  [P(k)+\bar{n}^{-1}]^2\, j_0(ks)^2,
\label{xiGerr}
\end{equation}
where $V$ is the volume of the survey (see Table~1), $P(k)$ the galaxy power spectrum, and $j_0$ the spherical Bessel 
function of zeroth order.  We first test this formula against the variance derived from the mock catalogs for both Dim 
and Bright samples, and find that
at $50<s<100$\hmpcii~the variance is consistent with Eq.~\ref{xiGerr}. At smaller scales non-Gaussian contributions to 
the 
errors quickly make Eq.~(\ref{xiGerr}) an understimate of the uncertainties, whereas at scales $s>100$\hmpcii~contributions 
to Eq.~(\ref{xiGerr}) from edge effects (not included there) become important, e.g. making the total error about 20\% larger 
than given by Eq.~(\ref{xiGerr}) at 300\hmpcii. Since edge effects depend only on the geometry and density of the sample, we 
can extract their value from this comparison and apply it to the data, which has the same characteristics (but different 
two-point function).  

From the measured two-point function in the data we use Eq.~(\ref{xiGerr}) plus the edge effects variance to compute Gaussian errors for a model that matches the observed $\xi$. We find uncertainties that are very close to those in the mock catalogs at $s>150$\hmpcii, the main reason being that discreteness contributions are significant. This suggests that had we changed cosmology and HOD to give a better fit to $\xi$ at largest scales, we would have gotten very similar uncertainties, and thus similar $\sim 2\sigma$ deviations away from zero-crossing for the Bright sample.




\clearpage

\onecolumn

\clearpage

\clearpage

\clearpage

\clearpage

\clearpage

\clearpage

\clearpage

\clearpage

\clearpage

\clearpage

\clearpage

\clearpage

\clearpage

\clearpage


\clearpage

\clearpage

\clearpage

\clearpage

\clearpage


\clearpage


\begin{deluxetable}{ccccccccccc}
\rotate
\tabletypesize{\footnotesize}
\small
\tablewidth{551pt}
\tablecaption{SDSS LRG Samples}{\label{sdsssamples}}
\tablehead{
\colhead{Sample} & 
\colhead{\# of LRGs} &
\colhead{$z_{\mathrm{min}}$} &
\colhead{$z_{\mathrm{max}}$} & 
\colhead{$\avg{z}$}  &
\colhead{$M_{g,\mathrm{min}}$} & 
\colhead{$M_{g,\mathrm{max}}$} &
\colhead{$\avg{M_g}$} & 
\colhead{Area} & 
\colhead{Volume} & 
\colhead{Density} \\
& 
&
&
& 
&
&
&
&
\colhead{(deg$^{-2}$)} & 
\colhead{($h^{-3}$ Gpc$^{3}$)} & 
\colhead{($10^{-5}$ $h^3$ Mpc$^{-3}$)} \\
}
\startdata
DR3\tablenotemark{a} & $47,063$  & $0.16$    &    $0.47$    &  $0.327$  & $-23.2$  & $-21.2$  &  $-21.70$ & $3,807$  & $0.722$    &   $6.50$  \\      
DR7-Full  & $105,831$ & $0.16$    &    $0.47$    &  $0.324$  & $-23.2$
& $-21.2$  &  $-21.72$ & $7,908$  & $1.58$     &   $6.70$    \\    
\qvl      & $61,899$  & $0.16$    &    $0.36$    &  $0.278$  & $-23.2$
& $-21.2$  &  $-21.65$ & $7,189$  & $0.66$     &   $9.40$    \\    
\vl        & $30,272$  & $0.16$    &    $0.44$    &  $0.338$  & $-23.2$  & $-21.8$  &  $-22.02$ & $7,189$  & $1.19$     &   $2.54$    \\    
\vlii-Near  & $16,473$  & $0.16$    &    $0.36$    &  $0.284$  & $-23.2$  & $-21.8$  &  $-22.01$ & $7,189$  & $0.66$     &   $2.50$    \\    
\vlii-Far   & $13,799$  & $0.36$    &    $0.44$    &  $0.402$  & $-23.2$  & $-21.8$  &  $-22.03$ & $7,189$  & $0.53$     &   $2.60$    \\    
\vll      & $32,861$  & $0.16$    &    $0.36$    &  $0.282$  & $-22.6$  & $-21.6$  &  $-21.85$ & $7,189$  & $0.66$     &   $5.00$    \\    
\enddata
\tablenotetext{a}{Calculations of DR3 performed using $\Omega_{M0}=0.3$. }

\end{deluxetable}
\clearpage

\end{document}